\documentclass[aps,pre,showpacs,raggedbottom, 
amssymb,twocolumn,groupedaddress]{revtex4-1}
\usepackage{graphicx}
\usepackage{amsmath}
\usepackage{amsfonts}
\usepackage{amssymb}
\usepackage{epsfig,libertine}
\usepackage[libertine]{newtxmath}
\usepackage{color}
\usepackage{dsfont, hyperref, xcolor}
\usepackage{comment}
\usepackage{enumerate}

\newcommand{\abs}[1]{\left\vert#1\right\vert}

\newcommand{\bra}[1]{\langle#1\vert}
\newcommand{\ket}[1]{\vert#1\rangle}

\newcommand{\lland}{\;\;\textrm{and}\;\;}
\newcommand{\llor}{\;\;\textrm{or}\;\;}

\newcommand{\be}{\begin{equation}}
\newcommand{\ee}{\end{equation}}

\begin{document}

\title{Topological phases in the presence of disorder and longer-range couplings}

\author{Gianluca Francica, Edoardo Maria Tiburzi, Luca Dell'Anna}
\address{Dipartimento di Fisica e Astronomia ``G. Galilei'', Universit\`{a} degli Studi di Padova, via Marzolo 8, 35131 Padova, Italy}

\date{\today}

\begin{abstract}
We study the combined effects of disorder and range of the couplings on the phase diagram of one-dimensional topological superconductors. We consider an extended version of the Kitaev chain where hopping and pairing terms couple many sites. 
Deriving the conditions for the existence of Majorana zero modes, we show that either the range and the on-site disorder can greatly enhance the topological phases characterized by the appearance of one or many Majorana modes localized at the edges. We consider both a discrete and a continuous disorder distribution. Moreover we discuss the role of correlated disorder which might further widen the topological regions. 
Finally we show that in the purely long-range regime and in the presence of disorder, the spatial decay of the edge modes 
remains either algebraic or exponential, with eventually a modified localization length, as in the absence of disorder. 
\end{abstract}

\maketitle

\section{Introduction}
Topological materials can exhibit boundary modes, topologically protected if the symmetries of the system are not broken~\cite{Schnyder08,kitaev09}. 
A so-called bulk-boundary correspondence relates the existence of these boundary modes to the topology of the quantum phase which is fully characterized by bulk topological invariants.  
The Kitaev one-dimensional chain~\cite{kitaev03} is a paradigmatic model exhibiting a topologically non-trivial quantum phase with two Majorana zero modes (MZMs) localized at the edges. 
Experimental realizations of such topological superconductors employ 
{\color{black}one-dimensional arrays of magnetic impurities \cite{perge,pawlak,ruby} or semiconducting nanowires \cite{Sau10,Oreg10,Alicea10,Mourik12} on top of a conventional superconducting substrate}.
MZMs are of some importance for achieving full-scale quantum computation. Actually, they were originally proposed with the aim to realize a quantum register immune from decoherence \cite{alicea,akhmerov}. 
Theoretically, this goal should be achievable due to the possibility of fault-tolerant quantum 
computation which can be obtained at the physical level, instead of using quantum error-correcting 
codes. The Kitaev chain is a way of constructing decoherence-protected degrees of freedom in one-dimensional systems. The question whether topological properties, exhibited by this model, are affected by longer-range couplings or randomness, all phenomena which might occur in real experimental setups, are therefore of some relevance.  
Several studies have been performed to understand the effects on the MZMs of long-range couplings \cite{pupillo, alecce17, patrick,lepori}, where bulk-boundary correspondence is still under investigation \cite{lepori,gong,jones}, the effects of 
{\color{black}disorder~\cite{Motrunich01,Brouwer11,degottardi13,cai13,degottardi,gergs16,giuliano,lieu18,Habibi18,monthus18,monthus218,hua19, Hui14} and interactions~\cite{gergs16,Lobos12,Thomale13,McGinley17,Kells18,Wouters18}, by considering several setups for experimental realizations \cite{BrouwerPRB11,Lutchyn11,Akhmerov11,Sau12,Liu12,Hui15,Sau13,Liu17,Haim19,pan221,cook12,zhang16,pan20,pan21,pan22,Ahn21,cole16,Awoga17,Liu18}. Fermionic chains have also been examined in the presence of next-nearest neighbor couplings~\cite{zhu15}, 
and long-range pairing with incommensurate potentials~\cite{cai17}. }
In this work, we aim at investigating the combined effect of disorder and longer-distance couplings on the topological phase diagrams, considering a chain of fermions which can host one or many couples of MZMs at the edges. In the absence of disorder, the effect of longer-range couplings has been investigated in Ref.~\cite{alecce17}. On the other hand, the effect of uniform disorder, for the Kitaev chain with nearest-nearest neighbor couplings, has been studied in Ref.~\cite{gergs16}. We will show, both numerically and analytically,  that the combined effect of disorder and range of the couplings is that of promoting the topological phases, increasing both their number and their extension, 
concluding that these elements, occurring in the real experimental realizations of the model, can even better stabilize the topological order. 
{\color{black} %
Finally we note that longer-range chains can be related to multichannel topological superconductors (see, e.g., Ref.~\cite{rieder12}), where the role of disorder has also been investigated \cite{rieder13,rieder14,Altland14,Pekerten17}. 
In particular, some positive effects of weak disorder have been originally observed in the case of multichannel~\cite{Haim19} and multiband~\cite{Adagideli14} topological superconductors.}

\section{Model}
To study the interplay between the interaction range and the strength of an on-site disorder we consider the extended version of the Kitaev chain, taking into account $r$ neighbor couplings in the hopping and pairing terms, as done in Ref.~\cite{alecce17}, with local random energies $\mu_j$.  This model is described by the following Hamiltonian 
\begin{eqnarray}\label{eq. Hami D}
\nonumber H&=&%
-\sum_{\ell=1}^r \sum_{j=1}^{L-\ell} \left(w_\ell  a_j^\dagger a_{j+\ell}-
 \Delta_\ell a_j a_{j+\ell}+H.c.\right)\,\\
&& +\sum_{j=1}^L\mu_j\left(a_j^\dagger a_j-\frac{1}{2}\right)%
\end{eqnarray}
where $a_j$ ($a^\dagger_j$) annihilates (creates) a fermion in the site $j$, %
$\mu_j$ is the space dependent chemical potential, $w_\ell$ is the hopping amplitude and $\Delta_\ell$ is the superconducting pairing.
We can define the following Majorana operators $c_{2j-1}=a_j + a^\dagger_j$ and $c_{2j}=-i (a_j - a_j^\dagger)$, such that $\{c_i,c_j\}=2\delta_{i,j}$. The Hamiltonian in Eq.~(\ref{eq. Hami D}) can be rewritten as 
\begin{eqnarray}\label{eq. Hami}
\nonumber H &=& -\frac{i}{2} \sum_{j=1}^L \mu_j c_{2j-1}c_{2j} + \frac{i}{2} \sum_{\ell=1}^r (w_\ell+\Delta_\ell) \sum_{j=1}^{L-\ell} c_{2j} c_{2j+2\ell-1}\\
 && -\frac{i}{2} \sum_{\ell=1}^r (w_\ell-\Delta_\ell) \sum_{j=1}^{L-\ell} c_{2j-1} c_{2j+2\ell}\,.
\end{eqnarray}
The model belongs to the BDI class and can exhibit a positive number of MZMs per edge. The case of a homogeneous chemical potential $\mu_j=\mu$ has been extensively studied in Ref.~\cite{alecce17}, where, to find the Majorana modes, 
 a general transfer matrix approach has been introduced.
 
{\color{black} A recent proposal for realizing a BDI topological superconductor from a AIII topological insulator is reported in Ref. \cite{He}. The first  experimental proposal \cite{Sau10,Oreg10,Alicea10}, instead, makes use of a quantum wire put in proximity of a s-wave superconductor in the presence of spin-orbit coupling and magnetic field which generate a topological superconductor of class D. In this latter case the BDI class can be reached in the limit of strong magnetic field compared to the spin-orbit coupling.}

\subsection{Transfer matrix}
\label{sec.tranfermatrix}
The transfer matrix approach can be easily generalized to the case of an inhomogeneous chemical potential. We will consider the presence of disorder, so that $\mu_j$ is randomly distributed. 
To determinate whether MZMs exist, we look for a MZM of the form $b_1=\sum_{j=1}^L \psi_j c_{2j}$. The condition $[H,b_1]=0$ is satisfied if 
\begin{equation}\label{eq. eig. 00}
\sum_{\ell=1}^r (w_\ell+\Delta_\ell)\psi_{i-\ell}-\sum_{\ell=1}^r (w_\ell-\Delta_\ell)\psi_{i+\ell}+\mu_i\psi_i =0\,,
\end{equation}
so that the mode is localized at one 
edge, after imposing $r$ boundary conditions ($\psi_i =0$ for $-r< i \le 0$ or  $L < i \le L+r$). 
One can find an analogous condition for the zero mode at the other 
edge, which reads 
\begin{equation}\label{eq. eig. 0}
\sum_{\ell=1}^r (w_\ell+\Delta_\ell)\phi_{i+\ell}-\sum_{\ell=1}^r (w_\ell-\Delta_\ell)\phi_{i-\ell}+\mu_i\phi_i =0
\end{equation}
 after requiring $[H,b_2]=0$, with $b_2=\sum_{j=1}^L \phi_j c_{2j-1}$, and 
 imposing $r$ boundary conditions ($\phi_i =0$ for $L < i \le L+r$ or $-r< i \le 0$). 
In particular, Eq.~\eqref{eq. eig. 0} can be expressed as 
\begin{equation}
\vec{\Phi}_{i+1}={\cal A}_i\vec\Phi_{i}
\end{equation} 
where %
$\vec \Phi_i = (\phi_{i+r-1},\phi_{i+r-2},\cdots,\phi_{i},\cdots,\phi_{i-r})^T$, and ${\cal A}_i$
is the transfer matrix 
\begin{equation}
{\cal A}_i = \left(
        \begin{array}{ccccc}
          t_1 & t_2 & \cdots & t_{2 r-1} & t_{2r} \\
          1 & 0 & \cdots & 0 & 0 \\
          0 & 1 & \cdots & 0 & 0\\
          \vdots & \vdots & \ddots  & \vdots & \vdots\\
          0 & 0 & \cdots & 1 & 0 \\
        \end{array}
      \right)\,,
\end{equation}
where $t_\ell=-\frac{\Delta_{r-\ell}+w_{r-\ell}}{\Delta_r+w_r}$ for $1\leq \ell <r$, $t_r=-\frac{\mu_i}{\Delta_r+w_r}$ and $t_\ell=\frac{\Delta_{\ell-r}-w_{\ell-r}}{\Delta_r+w_r}$ for $r<\ell \leq 2r$. 
By considering the eigenvalues of the matrix $\tilde {\cal A}_n$, defined as 
\begin{equation}
\label{tildeAcal}
\tilde {\cal A}_n = {\cal A}_n {\cal A}_{n-1}\cdots {\cal A}_2{\cal A}_1\,,
\end{equation}
we get a number $N=\max(v^>,v^<)-r$ of MZMs, where $v^>$ and $v^<$ are the numbers of eigenvalues $\lambda_n$ that diverge and tend to zero as $n\to \infty$ (see Ref.~\cite{alecce17} for details).

For $w_\ell=\Delta_\ell$, the condition expressed in Eq.~(\ref{eq. eig. 0}) reduces to 
$2 \sum_{\ell=1}^r w_\ell\phi_{i+\ell}+\mu_i\phi_i =0$,
which can be expressed as 
\begin{equation}
\vec\phi_{i+1}=A_i \vec\phi_{i}\,,
\end{equation}
where now %
$\vec \phi_i = (\phi_{i+r-1},\phi_{i+r-2},\cdots,\phi_{i})^T$, 
and $A_i$ is the transfer matrix
\begin{equation}
A_i = \left(
        \begin{array}{ccccc}
	t_1 & t_2 & \cdots & t_{r-1} & t_{r} \\
          1 & 0 & \cdots & 0 & 0 \\
          0 & 1 & \cdots & 0 & 0\\
          \vdots & \vdots & \ddots  & \vdots & \vdots\\
          0 & 0 & \cdots & 1 & 0 \\
        \end{array}
      \right)\,.
\end{equation}
where $t_\ell=-\frac{w_{r-\ell}}{w_r}$ for $\ell=1,\dots, r-1$, and $t_r=-\frac{\mu_i}{2w_r}$. 
Indeed, for $w_\ell=\Delta_\ell$, 
\begin{equation}
{\cal A}_i = \left(
        \begin{array}{cc}
        A_i& {0}\\
        \Sigma_+&\Sigma_-
        \end{array}
\right)
\end{equation}
where $\Sigma_+$ is a $r\times r$ matrix with all entries equal to zero except the right-top element equal to $1$, while $\Sigma_-$ is a $r\times r$ matrix with all entries equal to zero except $(r-1)$ elements along the low diagonal, below the main diagonal, which are equal to $1$. Actually $\Sigma_-$ does not play any role since it gives trivial identities. As a result 
\begin{equation}
\tilde {\cal A}_n = \left(
        \begin{array}{cc}
        \tilde A_n& {0}\\
       \dots&\Sigma_-^n
        \end{array}
\right)
\end{equation}
where analogously to Eq.~(\ref{tildeAcal}) we define
\begin{equation}
\label{tildeA}
\tilde {A}_n = {A}_n {A}_{n-1}\cdots {A}_2{A}_1\,.
\end{equation}
and for large distances, $n\ge r$, we have $\Sigma_-^n=0$. 

Finally, for an infinite range, $r\to \infty$, and an algebraic decay of the couplings, $w_\ell=w/\ell^\alpha$ and $\Delta_\ell=\Delta/\ell^\beta$, 
for $\alpha$ and $\beta>1$, 
we can have only 0 or 1 MZM, algebraically localized at each edge or exponentially localized for $w_\ell=\Delta_\ell$~\cite{francica22,jager20}.

\section{Nearest-neighbor couplings ($r=1$)}
The case of nearest neighbor coupling, $r=1$, has been investigated in Ref.~\cite{gergs16}. For $w\neq \Delta$ we get the transfer matrix
\begin{equation}
{\cal A}_i = \left(
        \begin{array}{cc}
          -\frac{\mu_i}{\Delta + w} & \frac{\Delta-w}{\Delta+w} \\
          1 & 0  \\
        \end{array}
      \right)\,,
\end{equation}
thus there is 1 MZM per edge if $v^>=2$ or $v^<=2$. 

For the case $w_1=\Delta_1$, from Eq.~\eqref{eq. eig. 0} we get 
$\phi_{i+1}=m_i\phi_i$, from which
\begin{equation}
\label{phin1}
\phi_{n+1} = \phi_1 \prod_{i=1}^n m_i,
\end{equation}
where we defined 
\begin{equation}
m_i \equiv -\frac{\mu_i}{2 w}\,.
\end{equation}
The only non-zero eigenvalue of the transfer matrix product is $\lambda=\prod_{i=1}^n m_i$, which is equal to $\tilde A_n$.
    Let us now rewrite the absolute value of $\phi_{n+1}$ as follows
    \begin{equation}
    \label{eq:3_eigen_r=1}
        \left|\phi_1\right|\prod_{i=1}^n\left|m_i\right|=
        \left|\phi_1\right|\exp \left[\sum_{i=1}^n\ln \left|m_i\right|\right]=
        \left|\phi_1\right|e^{n\langle \ln \left|m_i\right|\rangle}.
    \end{equation}

    In the thermodynamic limit, the topological phase is determined by the behavior of the mean logarithm $\langle \ln \left|m_i\right|\rangle\equiv \sum_{i=1}^n \ln \left|m_i\right|/n$, for large $n$, if this quantity is negative, the eigenvalue $\lambda$ vanishes and a topological phase is allowed, otherwise the phase is trivial, i.e. there is one MZM per edge if %
\begin{equation}
\label{eq. r1}
\langle \ln |m_i|\rangle < 0\,.
\end{equation}
This condition can be fixed for any probability distribution of the random variables. 

\begin{figure}
[t!]
\centering
\includegraphics[width=0.495\columnwidth]{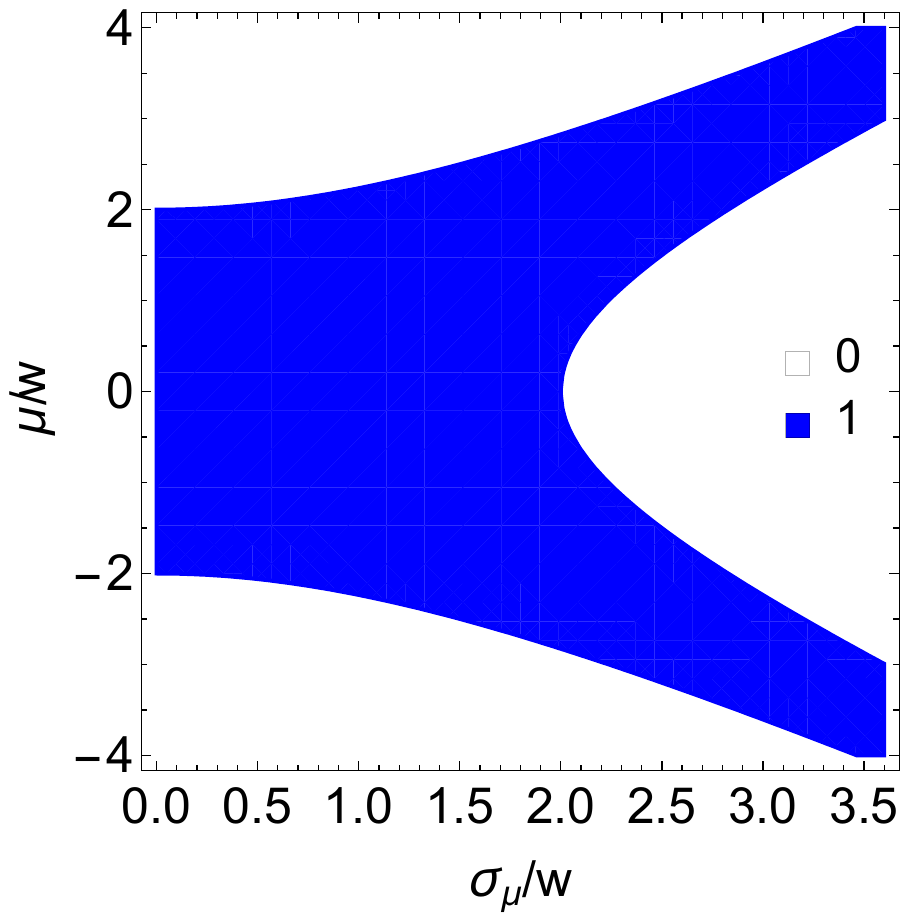}
\includegraphics[width=0.495\columnwidth]{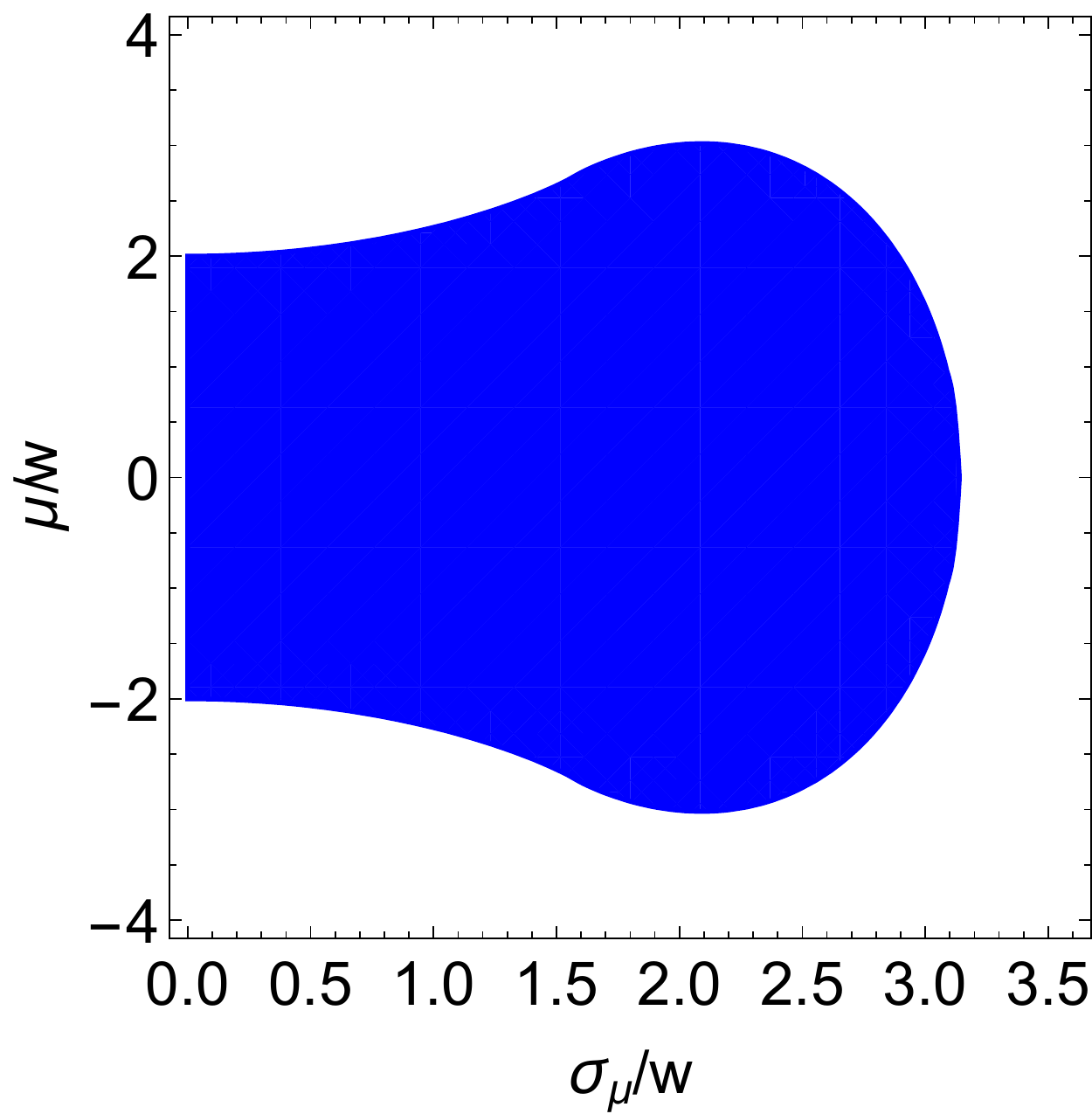}
\caption{Topological phase diagrams $\sigma_\mu$-$\mu$, for $r=1$ and $w=\Delta$, where $\mu$ and $\sigma_\mu^2$ are, respectively, the mean value and the variance of the distribution of the random variables $\mu_i$. The blue regions are characterized by $1$ MZM, while the white regions are trivial, with $0$ MZMs, for two different distributions of the disorder: bimodal (left) and uniform box (right) distributions.}
\label{fig:topo r=1}
\end{figure}

For instance, for a bimodal distribution, for which $m_i$ can have only two values, $m_+=m+\epsilon$ and $m_-=m-\epsilon$, with the same probability, so that the probability distribution is $\frac{1}{2}\delta(m-m_+)+\frac{1}{2}\delta(m-m_-)$.
The topological phase is, then, bounded by the solution of the following equation
\begin{equation}
\label{eq. bim}
\frac{1}{2} \left(
        \ln \left|m+\epsilon\right|+\ln\left|m-\epsilon\right|\right)=0\,.
\end{equation}
Strikingly, for this special situation the topological phase extends indefinitely, even for large disorder, as far as $m\simeq \mp \epsilon$, namely for $\mu\simeq \pm\sigma_\mu$, where $\mu$ is the mean value and $\sigma_\mu^2$ the variance of the distribution of $\mu_i$. At the band center, $\mu=0$, the topological phase closes at $\epsilon=1$, i.e. $\sigma_\mu =2w$ (see Fig.~\ref{fig:topo r=1}).

Another example is the box distribution \cite{gergs16}, when $m_i$ are uniformly distributed in the interval $[{m}-\epsilon,{m}+\epsilon]$. We get a topological phase boundary solving the equation $\langle \ln |m_i|\rangle = 0$ which, in the continuum, reads $\frac{1}{2\epsilon} \int_{{m}-\epsilon}^{{m}+\epsilon}\text{d}m'  \ln \left|m'\right|=0$, or more explicitly, %
\begin{equation}
\frac{m+\epsilon}{2\epsilon}\ln \left|m+\epsilon\right|-\frac{m-\epsilon}{2\epsilon}\ln \left|m-\epsilon\right|-1=0\,.
\end{equation}
This analytical condition gives the boundary for the topological phase in terms of the mean chemical potential $\mu/w=2m$ and of the square-root of the variance $\sigma_\mu/w=2\epsilon/\sqrt{3}$, for a uniform distribution, reported in Fig.~\ref{fig:topo r=1}. 
The plot shows a peculiar behavior of the response of the system to the presence of disorder. As the strength of the disorder increases, the topological phase becomes wider and wider up to an optimal value beyond which it is suppressed and then disappears at $\sigma_\mu/w=2e/\sqrt{3}$. 
For $w\neq \Delta$ the behavior of the topological phase is qualitatively the same, is simply more or less extended depending on whether $|\Delta|>|w|$ or viceversa. Also the normal distribution gives the same qualitative behavior. 

We plan to generalize this study for a longer range of the couplings, $r>1$, in order to investigate the combined effects of disorder and long-range couplings.

\section{Next-nearest-neighbors couplings ($r=2$)}
\label{Sec.NNN}

Let us now consider a range $r=2$, namely involving the nearest and next-nearest neighbor couplings.
For simplicity we will mainly focus on $w_\ell=\Delta_\ell$. The case of larger $r$ shares similar features of the case $r=2$, 
so we will also investigate the next-nearest-neighbor in more detail because it is simpler from an analytical point of view.

First let us consider the case with $w_1 = 0$ and $w_2\neq 0$. From Eq.~\eqref{eq. eig. 0} we get
\begin{equation}
\phi_{i+2}+\frac{\mu_i}{2w_2}\phi_i =0\,,
\end{equation}
from which we obtain two equations for $i_0=1,2$
\begin{equation}
\label{phin2}
\phi_{2n+i_0}=\prod_{i=0}^{n-1} \left(-\frac{\mu_{2i+i_0}}{2 w_2}\right)\phi_{i_0}\,,
\end{equation}
which generalizes Eq.~(\ref{phin1}). So similarly to what done for $r=1$, we have now two copies of the condition for getting one MZM
\begin{equation}\label{eq. con 2}
\left\langle \ln \left| \frac{\mu_{i}}{2 w_2}\right|\right\rangle_{i_0}  < 0\,,
\end{equation}
where we define the averages %
$\langle x_i \rangle_{i_0} =  2 \sum_{i=0}^{n-1} x_{2i + i_0}/n$, valid for large $n$.
If locally the distribution probability of the disorder is the same at any site, the two conditions in Eq.~\eqref{eq. con 2} (with $i_0=1$ and $2$) are equal, and we can get only zero or $2$ MZMs per edge. 

In contrast, things change drastically if also the coupling $w_{1}\neq 0$.  Let us start considering the case $w_1=w_2=w$, so that the transfer matrix reads
\begin{equation}
A_i=\left(
  \begin{array}{cc}
    -1 & m_i \\
    1 & 0 \\
  \end{array}
\right)\,.
\end{equation}
To determinate the eigenvalues $\lambda_n$ of $\tilde A_n$, we consider the eigenvalue equation
\begin{equation}\label{eq. lambda 1}
\lambda_n^2 - \lambda_n \text{Tr}(\tilde A_n) + \det(\tilde A_n) = 0\,.
\end{equation}
Using the property of the determinant of a matrix product we have simply 
\begin{equation}
\det(\tilde A_n) =\prod_{i=1}^n \det(A_i) =(-1)^n\prod_{i=1}^n m_i
\end{equation}
and taking the absolute value we can write
\begin{equation}\label{eq. det}
|\det(\tilde A_n)| = \prod_{i=1}^n |m_i| = e^{n \langle \ln| m_i| \rangle}\,\,
\end{equation}
as $n\rightarrow \infty$.
The calculation of the trace is slightly more involved. 
As shown in Appendix~\ref{app. trace}, 
we get
\begin{equation}\label{eq. trace}
|\text{Tr}(\tilde A_n)| \simeq e^{n \langle \ln|R_i| \rangle}\,,
\end{equation}
as $n\to \infty$, where we define the random continued fraction
\begin{equation}
\label{continued}
R_{i} = 1 + \cfrac{m_i}{1+\cfrac{m_{i-1}}{1+\cfrac{m_{i-2}}{1+\ddots}}}\,,
\end{equation}
which is the solution of the following recursive equation
\begin{equation}
\label{recursive}
R_i = 1+\frac{m_i}{R_{i-1}}
\end{equation}
We note that the mean values $\langle \ln|m_i|\rangle$ and $\langle \ln|R_i| \rangle$ can be evaluated by using the probability distributions of the random variables $m_i$ and of the convergents $R_i$. Unlike the determinant, the trace depends on disorder in a non-trivial way. As shown in Appendix~\ref{app. trace} the trace reads
\begin{equation}
\label{TrAn}
\text{Tr}(\tilde A_n)\approx 1+ n\langle m_i\rangle + \frac{n^2}{2}\langle m_i\rangle^2- \frac{n}{2}\langle m_i^2\rangle - n\, \langle m_i m_{i-1}\rangle  + \cdots\,,
\end{equation}
where we neglected terms depending on higher moments and correlations. Clearly, the topological quantum phases can depend on both moments (e.g., $\langle m_i^2\rangle$) and correlations (e.g., $\langle m_i m_{i-1}\rangle$) of the random variables. As a result 
we expect that correlated and uncorrelated disorder can affect differently the topological phases. 
This behavior has to be contrasted to the case with $w_{1} = 0$ %
where correlations do not play any role.

In order to derive the conditions for the existence of MZMs, for $w_\ell= \Delta_\ell$, we have to analyze the eigenvalue equation, Eq.~\eqref{eq. lambda 1}. 
If $\det(\tilde A_n) \to 0$ or $\det(\tilde A_n)/\text{Tr}(\tilde A_n) \to 0$,  from Eq.~\eqref{eq. lambda 1} we get
$\lambda_n(\lambda_n - \text{Tr}(\tilde A_n)) \to 0$ or $\lambda_n(\lambda_n/\text{Tr}(\tilde A_n) - 1) \to 0$, thus in both cases there is at least one eigenvalue %
that tends to zero, so that there is at least 1 MZM per edge. 
If $\text{Tr}(\tilde A_n) \to 0$ and $\det(\tilde A_n)\to 0$, for $n\to\infty$, from Eq.~\eqref{eq. lambda 1} we get $\lambda_n \to 0$, thus both the two eigenvalues %
go to zero, so that there are $2$ MZMs per edge. 
In summary, we get the following conditions
\begin{eqnarray}
\label{eq. c1}
&&\det(\tilde A_{n}) \to 0 \llor \frac{\det(\tilde A_{n})}{\text{Tr}(\tilde A_{n})} \to 0 \Rightarrow \exists\,\text{1 MZM}\,,\\
\label{eq. c2}
&&\det(\tilde A_{n}) \to 0 \lland  \text{Tr}(\tilde A_{n}) \to 0 \Rightarrow \exists\,\text{2 MZMs}\,.
\end{eqnarray}
For $w_1=w_2=w$, using Eqs.~\eqref{eq. det} and~\eqref{eq. trace}, we get
\begin{eqnarray}
&&\label{eq. c1-f} \langle \ln|m_i| \rangle <0 \llor  \langle \ln|m_i| \rangle < \langle \ln|R_i| \rangle  \Rightarrow \exists\,\text{1 MZM}\,,\\
&&\label{eq. c2-f}\langle \ln|m_i| \rangle <0 \lland \langle \ln|R_i| \rangle <0  \; \Rightarrow \exists\,\text{2 MZMs}\,.
\end{eqnarray}
We now proceed studying these conditions for the homogeneous case and then applying them to the cases of an uncorrelated and a correlated disorder.

\subsection{Homogeneous case}
Let us warm up considering first the clean system, namely in the absence of disorder, always in the simplest case where $w_1=w_2=\Delta_1=\Delta_2\equiv w$. In this case $m_i=m=-\mu/(2w)$ for any $i$, so that we get
\begin{eqnarray}
&&\det(\tilde A_{n})=(-1)^n m^n\,,\\
&&\text{Tr}(\tilde A_{n})=\frac{(-1)^n}{2^n}\left[\left(1+\sqrt{1+4 m}\right)^n+\left(1-\sqrt{1+4 m}\right)^n\right].\;
\end{eqnarray}
For $n\rightarrow \infty$ we get $\frac{\det(\tilde A_{n})}{\text{Tr}(\tilde A_{n})} \to 0$ for $m\in (-1,2)$ so that, according to Eq.~(\ref{eq. c1}), we have at least $1$ MZM per edge for $\mu/w\in (-4,2)$. Moreover, $\text{Tr}(\tilde A_{n}) \to 0$ for $m\in (-1,0)$, so that, using Eq.~(\ref{eq. c2}), we get $2$ MZMs for $\mu/w\in (0,2)$. 
These results are in perfect agreement with what reported in Ref.~\cite{alecce17}.

Let us redo the calculation using Eqs.~(\ref{eq. c1-f}), (\ref{eq. c2-f}), introducing the distribution $\rho(R_i)$ for a continued fraction, which will be useful also in the presence of disorder. 
For $m>-1/4$ we get $R_n\to \bar R = \frac{1}{2}(1 + \sqrt{1+4m})$ for $n\to \infty$, so that $\rho(R_i) = \delta(R_i - \bar R)$. To determinate the distribution $\rho(x)$ for $m<-1/4$, we consider the Frobenius-Perron equation (see, e.g., Ref.~\cite{ott93})
\begin{equation}
\rho(x)=\int \rho(y) \delta(x-F(y))dy\,,
\end{equation}
where $F(y)=1+m/y$ (see Eq.~(\ref{recursive})) from which we get %
\begin{equation}\label{eq distr 1}
\rho(x) = \rho\left(\frac{m}{x-1}\right)\frac{|m|}{(x-1)^2}\,.
\end{equation}
For $m<-1/4$ a solution is given by the Lorentzian function
\begin{equation}
\rho(x) = \frac{1}{\pi}\frac{\gamma}{\left(x-\frac{1}{2}\right)^2+\gamma^2}
\end{equation}
with $\gamma=\sqrt{|4m+1|}/2$. Using this distribution we get 
\begin{equation}
\langle \ln |R_i|\rangle =  \int \rho(x) \ln|x|dx = \frac{1}{2}\ln|m|
\end{equation}
for $m<-1/4$, and 
\begin{equation}
\langle \ln |R_i|\rangle =\ln|\bar R|=\ln\left|(1 + \sqrt{1+4m})/2\right|
\end{equation}
 for $m>-1/4$. It is easy to see that, from Eq.~\eqref{eq. c2-f}, there are 2 MZMs per edge if $m\in (-1,0)$, and, from Eq.~\eqref{eq. c1-f}, there is 1 MZM per edge if $m\in(-1,2)$, in agreement with Ref.~\cite{alecce17}.

\subsection{Uncorrelated disorder}
Let us now consider uncorrelated random variables $m_i$.  For simplicity, we will consider a bimodal disorder, so that $m_i$ can be equal to $m_+=m+\epsilon$ and $m_-=m-\epsilon$, with the same probability.  As shown in Ref.~\cite{loreto96}, the distribution $\rho(R_i)$ is the solution of
\begin{equation}
\rho(x)= \frac{1}{2}\int \rho(y) \left[\delta(x-F(y;m_+))+ \delta(x-F(y;m_-))\right]dy
\end{equation}
where $F(y;m_{\pm})$ is the map $F(x)$ with $m_i=m_\pm$, %
from which we get the equation
\begin{equation}\label{distr diso}
\rho(x) =  \frac{1}{2} \rho\left(\frac{m_+}{x-1}\right)\frac{|m_+|}{(x-1)^2} + \frac{1}{2}\rho\left(\frac{m_-}{x-1}\right)\frac{|m_-|}{(x-1)^2}\,.
\end{equation}
We look for a non-trivial solution of this equation. For $m<-1/4$,  we get
\begin{equation}
\rho(x) = \frac{1}{\pi} \frac{\gamma}{\left(x-\frac{1}{2}\right)^2+\gamma^2} + \eta(x)\,,
\end{equation}
where $\gamma=\sqrt{\abs{4 m+1}}/2$, and $\eta(x) \sim \epsilon^2$ since Eq.~\eqref{distr diso} is satisfied for $\eta(x)=0$ up to terms of the order $\epsilon^2$. In particular, $\eta(x)$ is solution of
\begin{equation}
\eta(x) = g(x) + \eta\left(\frac{m}{x-1}\right) \frac{|m|}{(x-1)^2} \,,
\end{equation}
where
\begin{equation}
g(x) = - \frac{\sqrt{\abs{1+4 m}}(m^2+3m(x-1)^2-(x-1)^3)}{2\pi m^2(m+x-x^2)^3} \epsilon^2+ O(\epsilon^3)\,.
\end{equation}
Since we are interested only on the mean value $\langle \ln|R_i|\rangle$, we can use the following equations%
\begin{eqnarray*}
&&\int \eta(x) \ln|x|dx =- \int \big(\eta(x)-g(x)\big) \ln|x-1|dx\\
&&\int \eta(x) \ln|x| dx = \frac{1}{2} \int \big(\eta(x) \ln|x+m| dx + g(x) \ln|x|\big) dx\,,
\end{eqnarray*}
which can be easily combined for $m=-1$, getting
\begin{equation}
\int \eta(x) \ln|x| dx = -\frac{\epsilon^2}{12} + O(\epsilon^3)\,.
\end{equation}
The effect of the disorder on widening or shrinking the boundary of the topological phase can be understood by considering the critical points in the absence of disorder. For the topological phase with 2 MZMs, we have the boundaries at $m=-1$ and $m=0$ in the absence of disorder. \\
For $m=-1$,  we get
\begin{equation}
\langle \ln|R_i| \rangle =-\frac{\epsilon^2}{12} + O(\epsilon^3)\,.
\end{equation}
On the other hand, at $m=0$, we have that
\begin{equation*}
\rho(x) = \frac{1}{4}\big(\delta(x-R_+) + \delta(x-R_-)+\delta(x-R'_+) + \delta(x-R'_-)\big) + O(\epsilon^3)
\end{equation*}
is the solution of Eq.~\eqref{distr diso}, 
with $R_{\pm} = 1\pm \epsilon-\epsilon^2 + O(\epsilon^3)$, $R'_{\pm} = 1\pm \epsilon+\epsilon^2 + O(\epsilon^3)$.
As a result we get $\langle \ln |R_i|\rangle = \frac{1}{4} \left( \ln|R_+| + \ln|R_-| + \ln|R'_+| + \ln|R'_-|\right)$.\\
For $m=0$ we have, therefore,
\begin{equation}
\langle \ln |R_i|\rangle = - \frac{\epsilon^2}{2} + O(\epsilon^3)\,.
\end{equation}
On both boundaries $m=-1$ and $m=0$, we have $\langle \ln|R_i| \rangle <0$, and since also $\langle \ln|m_i|\rangle$ is negative, from Eq.~\eqref{eq. c2-f} we find that the topological region with 2 MZMs widens on both sides switching on a small disorder $\epsilon$.

On the other hand, for the topological phase with 1 MZM, we have the boundaries at $m=-1$ and $m=2$ in the absence of disorder.
At the point $m=-1$ we get
\begin{equation}
\langle \ln|m_i| \rangle =  -\frac{\epsilon^2}{2}+ O(\epsilon^3)\,.
\end{equation}
At $m=2$, we get that, for small $\epsilon$, $\rho(R_i)$ is uniform and equal to $1/(2\epsilon)$ for $R_i\in(2-\epsilon,2+\epsilon)$ and zero otherwise, so that $\langle \ln|R_i| \rangle\simeq \ln(2)-\epsilon^2/24$. Moreover close to $m=2$, for small $\epsilon$, from Eq.~(\ref{eq. bim}) we get $\langle \ln|m_i| \rangle\simeq \ln(2)-\epsilon^2/8$, so that 
\begin{equation}
\langle \ln|m_i| \rangle -\langle \ln|R_i| \rangle =  -\frac{\epsilon^2}{8}+ \frac{\epsilon^2}{24} + O(\epsilon^3)\,.
\end{equation}
From Eq.~\eqref{eq. c1-f} we find that also the topological region with 1 MZM widens for small $\epsilon$. We expect this behavior occurs also for other distributions of the disorder, namely that the disorder increases the topological phase, 
since only the variance $\sigma_m^2$ of the random variables appears at the leading order. 

\begin{figure}
[t!]
\centering
\includegraphics[width=0.495\columnwidth]{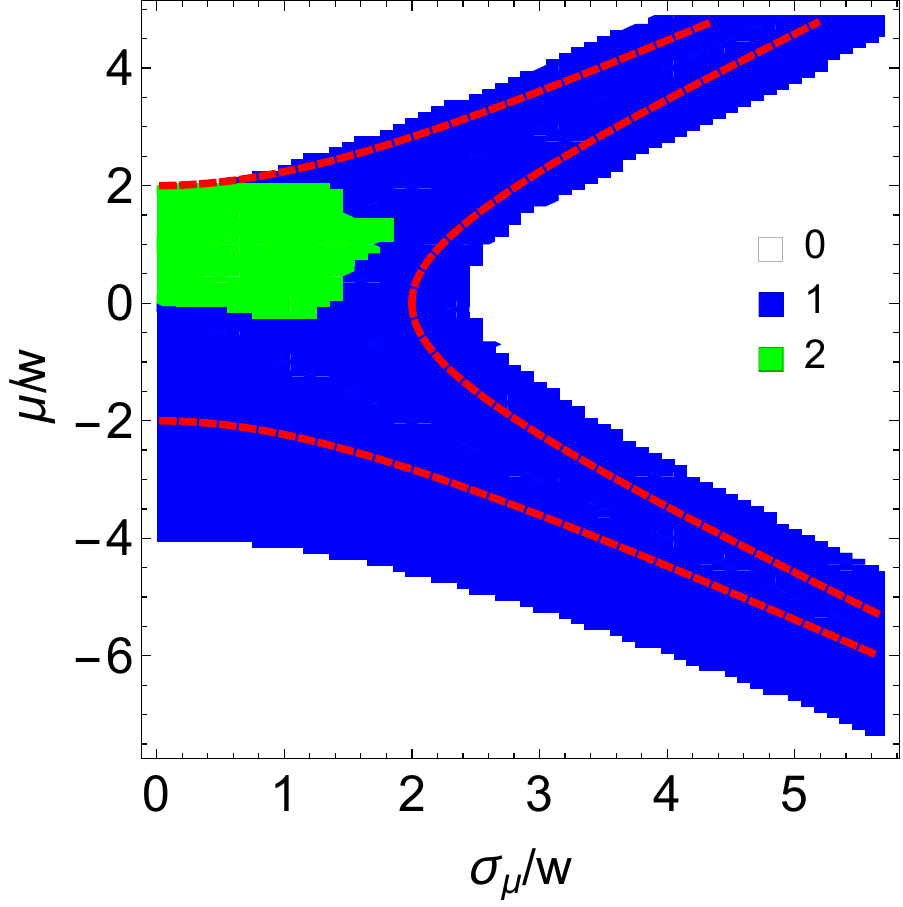} \includegraphics[width=0.495\columnwidth]{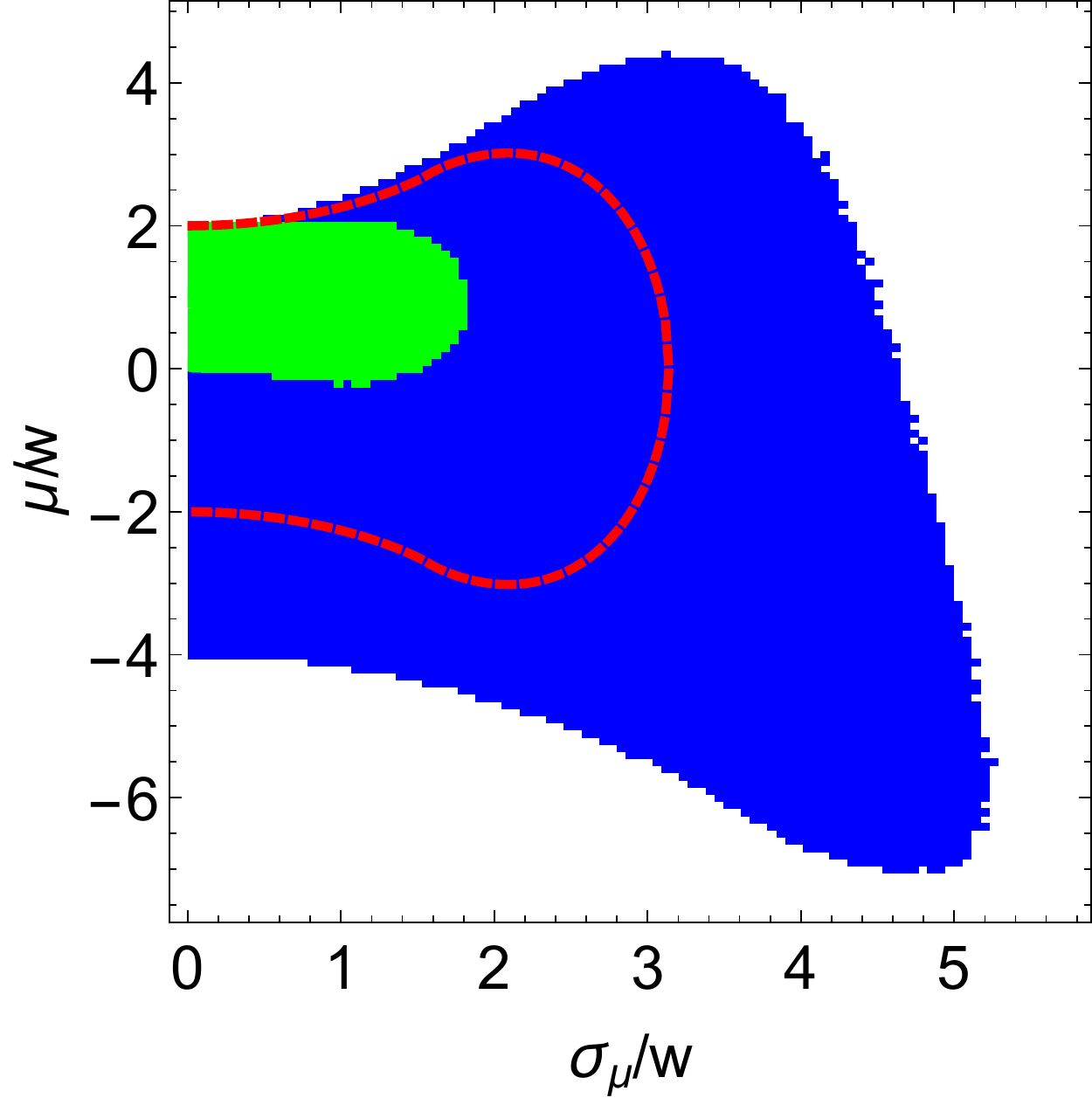}
\caption{
Topological phase diagrams $\sigma_\mu$-$\mu$, for next-nearest neighbor couplings, $r=2$, and $w_1=w_2=\Delta_1=\Delta_2$, obtained using the condition reported in Eqs.~\eqref{eq. c1-f} and~\eqref{eq. c2-f}, averaging over $n=10^5$ random variables. The diagrams are characterized by $0$ (white regions), $1$ (blue regions) and $2$ (green regions) MZMs per edge, for two different distributions of the disorder: bimodal (left) and uniform  box (right) distributions. For a better comparison with Fig.~\ref{fig:topo r=1}, the topological boundaries for $r=1$ are reported by red dashed lines. }
\label{fig:topo}
\end{figure}

Let us consider the case of large disorder. There are two different behaviors for the case of a continuous or discrete distribution. For the bimodal distribution, as shown in the left panel of Fig.~\ref{fig:topo}, the topological region with 1 MZM per edge survives for large disorder strength,  
near the resonances $m\approx \pm \epsilon$.
\begin{figure}
[t!]
\centering
\includegraphics[width=0.495\columnwidth]{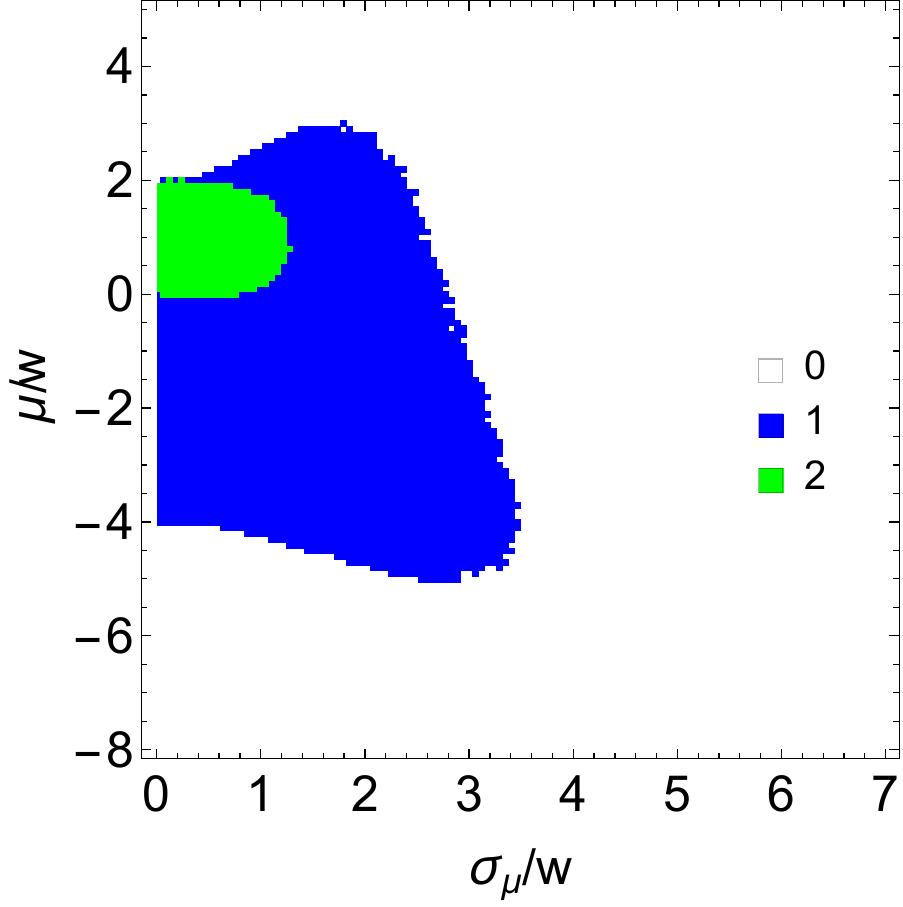} \includegraphics[width=0.495\columnwidth]{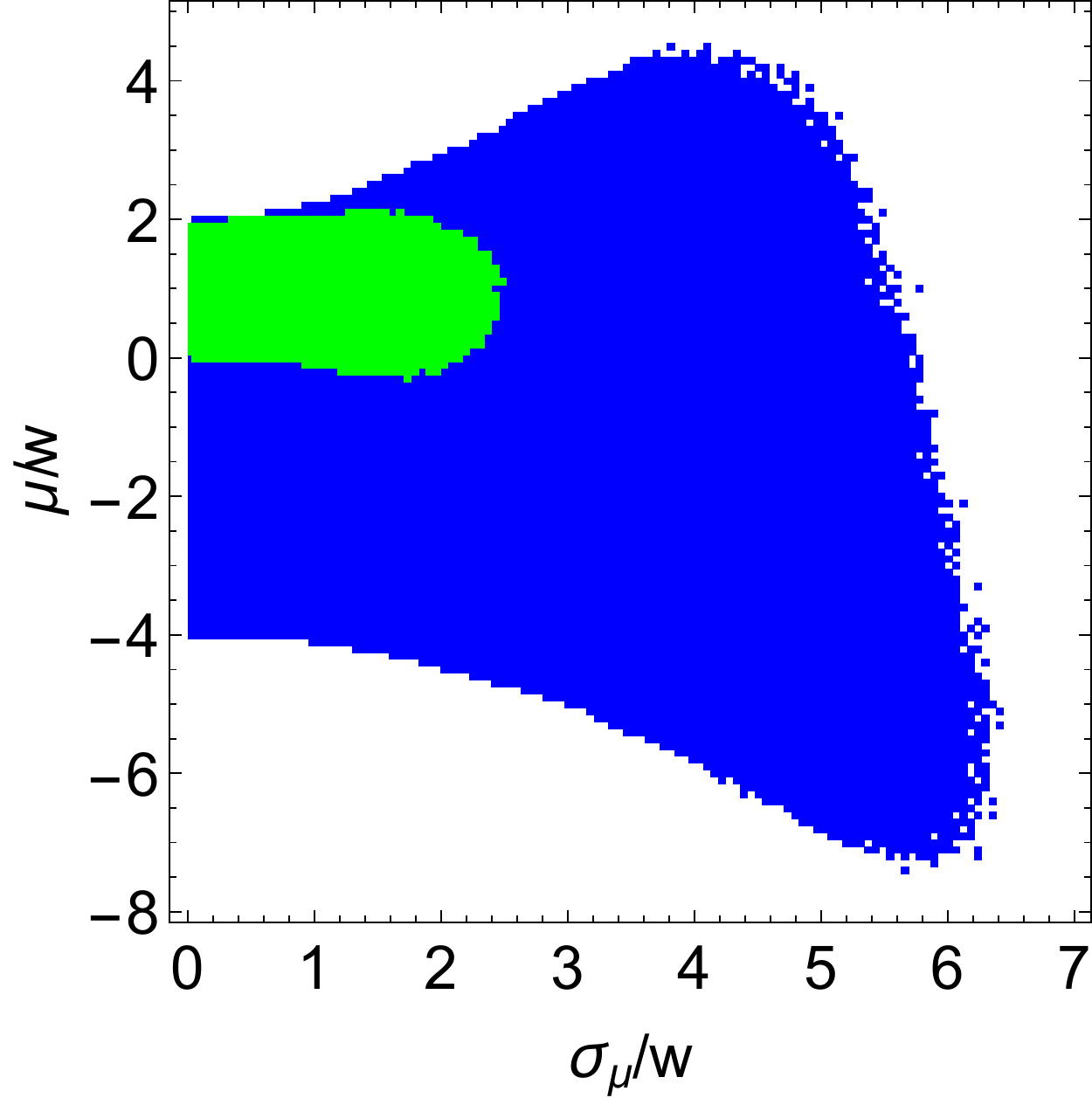}
\caption{
Topological phase diagrams $\sigma_\mu$-$\mu$, for next-nearest neighbor couplings, $r=2$, and for parameters $w=w_1=w_2$ and $\Delta=\Delta_1=\Delta_2$ but $\Delta\neq w$. 
 We consider a uniform disorder distribution and we average over $100$ disorder configurations for a chain with $L=200$ sites. 
 In the left panel $\Delta=w/2$ while in the right panel $\Delta=2w$. %
}
\label{fig:topo delta}
\end{figure}
This behavior can be explained by noticing that $\langle \ln|m_i|\rangle =1/2 (\ln|m+\epsilon|+\ln|m-\epsilon|)\rightarrow -\infty$ for $m\to \pm \epsilon$, so that the condition $\langle \ln|m_i|\rangle<0$ for having one MZM is always satisfied.
This behavior still exists for any discrete distribution of disorder with a finite number of values, while disappears when the values of the distribution become dense in a certain interval, since in this case $\langle \ln|m_i|\rangle$ cannot diverge and becomes positive for large disorder. 
For a box distribution, for instance, where $m_i$ are uniformly distributed in the interval $[m-\epsilon,m+\epsilon]$, the topological region is limited, as shown in the right panel of Fig.~\ref{fig:topo}. In any case, upon increasing the range of the couplings, we observe an increase of the whole topological phase, as well as the appearance of an additional phase characterized by two MZMs per edge. In Fig.~\ref{fig:topo}, in the topological phase diagram for $r=2$, also the boundaries for the case $r=1$ are reported (red dashed lines). 
Finally, we observe that for the case $w_\ell\neq\Delta_\ell$ we get essentially the same behavior. The only effect of having a larger or a smaller paring with respect to the hopping parameter is that the topological phase can be increased or suppressed mainly in the disorder strength direction as shown in Fig.~\ref{fig:topo delta}. In this latter case we determine the number of MZMs, by counting the number $v^<$ ($v^>$) of eigenvalues $\lambda_n$  of $\tilde {\cal A}_n$ such that $|\lambda_n|<1$ ($|\lambda_n|>1$), as explained in Sec.~\ref{sec.tranfermatrix}. 

\subsection{Correlated disorder}
\begin{figure}
[t!]
\centering
\includegraphics[width=0.45\columnwidth]{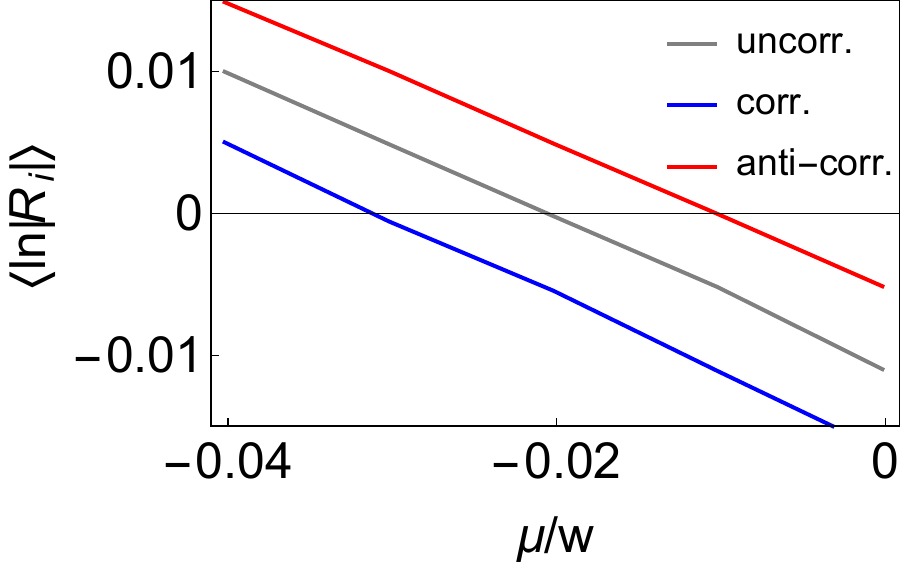}\hspace{0.1cm}
\includegraphics[width=0.49\columnwidth]{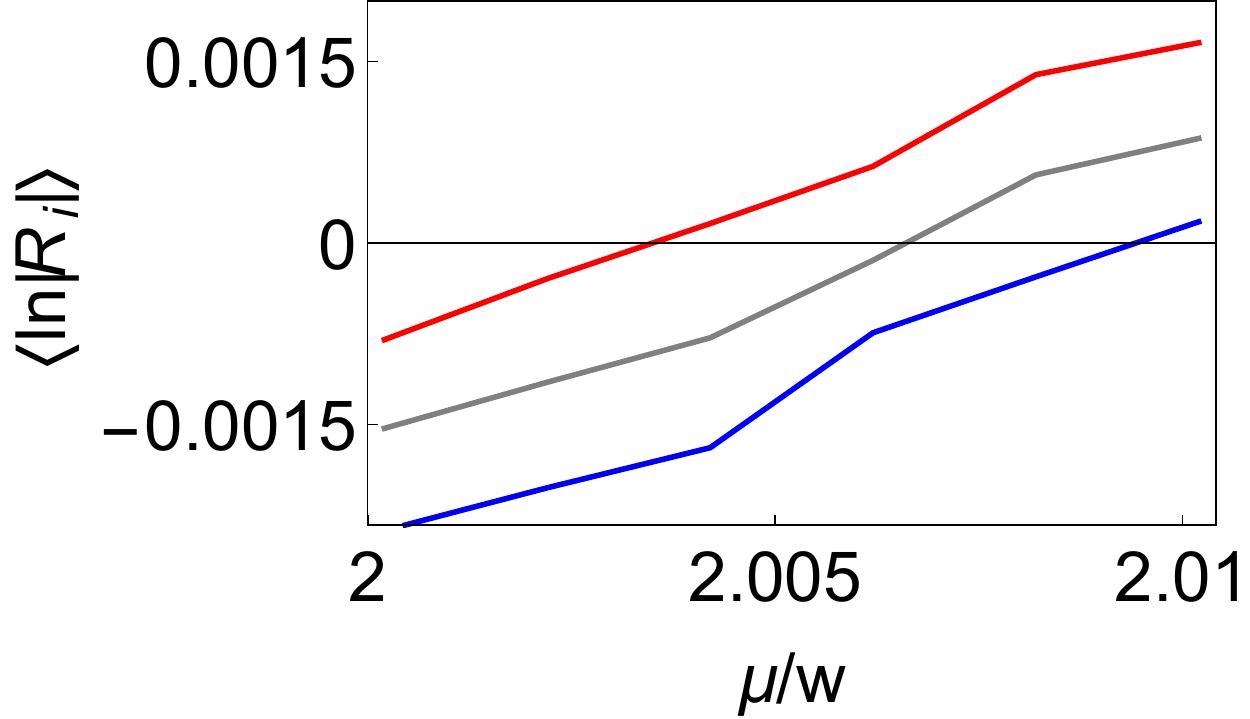}
\caption{ 
$\langle \ln |R_i|\rangle$ as a function of the mean value of the chemical potential $\mu$ for uncorrelated, correlated ($s=+1$) and anti-correlated ($s=-$) disorder, averaging over $n=10^6$ random variables,  
for $\epsilon=0.2$ (where $\sigma_\mu/w = 2\epsilon$). In the interval under consideration, $\langle \ln|m_i|\rangle < 0$ is verified, so that we get a topological phase with 2 MZMs per edge for $\langle \ln|R_i| \rangle<0$.}
\label{fig:topo2}
\end{figure}

\begin{figure}
[t!]
\centering
\includegraphics[width=0.49\columnwidth]{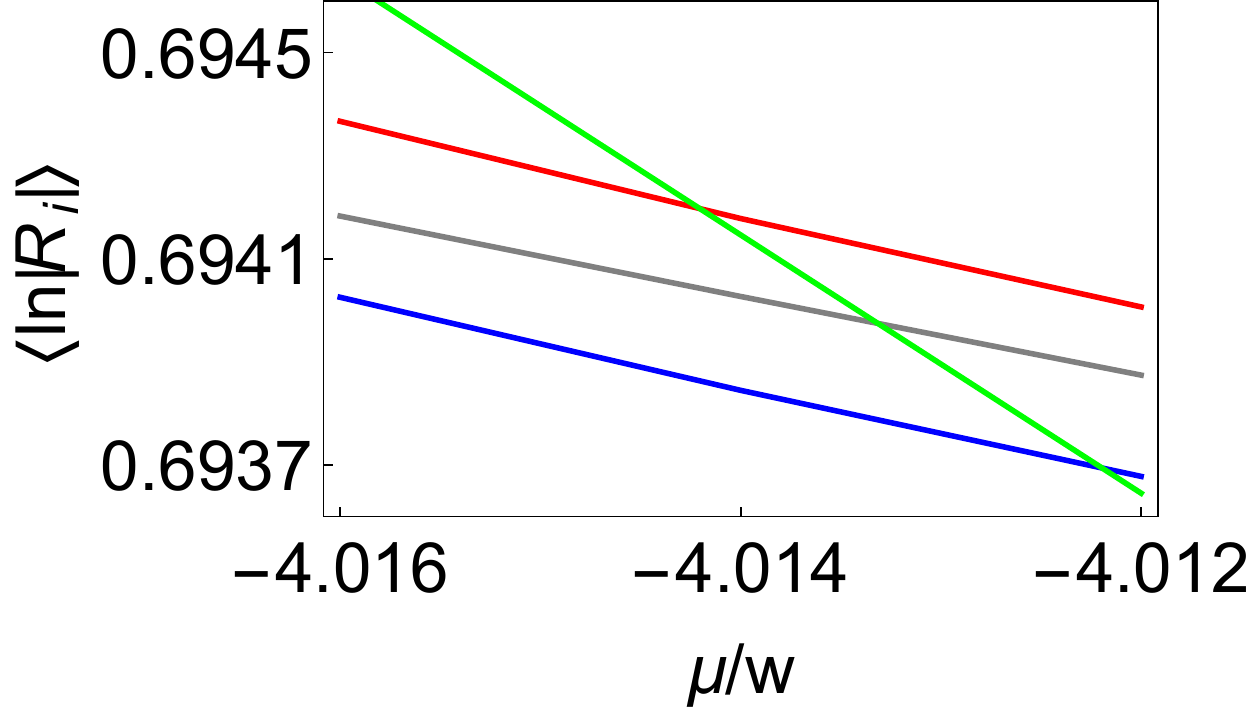}\hspace{0.1cm}
\includegraphics[width=0.47\columnwidth]{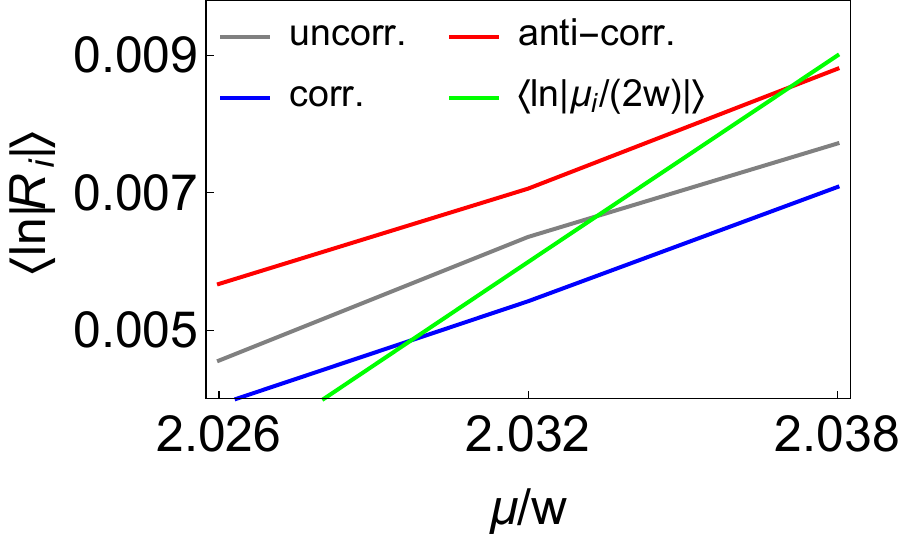}
\caption{$\langle \ln |R_i|\rangle$ as a function of $\mu$, 
for uncorrelated, correlated ($s=+1$) and anti-correlated ($\sigma=-1$) disorder, with $\epsilon=0.2$. The green line shows $\langle \ln|m_i|\rangle$ as a function of $\mu$.  
We get a topological phase with 1 MZM per edge for $\langle \ln|R_i| \rangle >\langle \ln|m_i| \rangle $.}
\label{fig:topo1}
\end{figure}

Let us now investigate the effects of a correlated disorder, namely when $\langle m_i m_j\rangle \neq \langle m_i \rangle \langle m_j\rangle$ for some $i$ and $j$. A way to introduce a correlation in the disorder is the following. Let us consider
\begin{equation}
m_i = m + \epsilon x_i
\end{equation}
The random variables $x_i$ can be chosen conveniently, in turn written in terms of other two random variables $y_i$ and $z_i$, in order to design two kinds of disorder:
\begin{enumerate}
\item uncorrelated disorder, such that $x_{i} = (y_{i} + z_{i})/2$,
\item correlated disorder, such that $x_{2i-1}= (y_{2i-1}+ z_{2i-1})/2$ and $x_{2i}=s\,(y_{2i-1}+ y_{2i})/2$, where $s=\pm 1$,
\end{enumerate}
where $z_i$ and $y_i$ are uncorrelated random variables which take values $-1$ or $1$ with equal probability. 
For both kinds of disorder, locally at any site $i$ we get the same distribution $p(m_i)=\frac{1}{2}p_{bi}(m_i) +\frac{1}{2} \delta(m_i-m)$, where $p_{bi}(m_i)=\frac{1}{2} \delta(m_i-m-\epsilon)+\frac{1}{2} \delta(m_i-m+\epsilon)$ is the bimodal one. As a result, 
$\langle \ln |m_i|\rangle = 1/2 \ln |m|+(\ln |m+\epsilon|+ \ln |m-\epsilon|)/4< 0$, i.e. the first condition in Eqs.~\eqref{eq. c1-f} and~\eqref{eq. c2-f}, is the same for both kinds of disorder. In contrast, the second condition in Eq.~\eqref{eq. c1-f} for one MZM, $\langle \ln|m_i|\rangle<\langle \ln|R_i|\rangle$,  and the second condition in Eq.~\eqref{eq. c2-f} for two MZMs, $\langle \ln |R_i|\rangle < 0$, can be different for the two kinds of disorder. As shown in Figs.~\ref{fig:topo2} and~\ref{fig:topo1}, correlations and anti-correlations have opposite effects.

In detail, for 2 MZMs, the presence of correlated disorder ($s=+1$) widens the topological phase, while the presence of anti-correlation ($s=-1$) shrinks it (see Fig.~\ref{fig:topo2}).
On the contrary, for 1 MZM, the presence of correlation ($s=+1$) shrinks  the topological phase, while anti-correlation ($s=-1$) widens it (see Fig.~\ref{fig:topo1}).

\subsection{Power-law coupling}
We conclude our investigation for the case $r=2$ by considering a power-law decay of the coupling parameters, $w_\ell = \Delta_\ell=w /\ell^\alpha$. In this case the transfer matrix reads
\begin{equation}
A_i=\left(
  \begin{array}{cc}
    -2^\alpha & 2^\alpha m_i \\
    1 & 0 \\
  \end{array}
\right)\,.
\end{equation}
In this case, for $n\to \infty$, the determinant in Eq.~(\ref{eq. det}) becomes 
\begin{equation}
|\det(\tilde A_n)| \simeq 2^{n\alpha} e^{n\langle\ln|m_i|\rangle}\,,
\end{equation}
while the trace in Eq.~(\ref{eq. trace}) is modified as follows%
\begin{equation}
|\text{Tr}(\tilde A_n)|\simeq 2^{n\alpha} e^{n \langle \ln|R^\alpha_i|\rangle}\,,
\end{equation}
where we define 
\begin{equation}
R^\alpha_i = 1 + \frac{m_i}{2^\alpha R^\alpha_{i-1}}\,,
\end{equation}
with $R^\alpha_1=1$, so that $R^0_i=R_i$.
This can be understood by noting that $\text{Tr}(\tilde A_n)$ for $\alpha>0$ is obtained from $\text{Tr}(\tilde A_n)$, with $\alpha=0$, by multiplying it by $2^{n\alpha}$ and by replacing $m_i$ with $m_i/2^\alpha$. The condition Eq.~(\ref{eq. c2-f}) for two MZMs, then, is modified as follows
\begin{equation}\label{eq. c2 alpha}
\langle \ln |R^\alpha_i|\rangle <-\alpha \ln 2 \lland \langle \ln |m_i|\rangle <-\alpha \ln 2   \Rightarrow \exists\,\text{2 MZMs}\,.
\end{equation}
We expect that $\langle \ln |R^\alpha_i|\rangle\geq -\ln 2$, where the equality holds for $m_i=-2^\alpha/4$, so that the condition $\langle \ln |R^\alpha_i|\rangle <-\alpha \ln 2$ can be satisfied only if $\alpha<1$.
For the homogeneous case, $m_i=m$, it is easy to see that the inequality $\langle \ln |R^\alpha_i|\rangle\geq -\ln 2$ is satisfied.
Furthermore, if $|m_i|\leq 2^\alpha/4$, the convergents are $R^\alpha_i\geq 1/2$ (see, e.g., Ref.~\cite{paydon42}), 
therefore also in this case $\langle \ln |R^\alpha_i|\rangle \geq -\ln2$. 
As a result we expect that, from Eq.~(\ref{eq. c2 alpha}) we can have 2 MZMs per edge only for $\alpha<1$.\\
On the other hand, concerning the topological phase with 1 MZM, Eq.~(\ref{eq. c1-f}) is modified getting the following condition
\begin{equation}\label{eq. c1 alpha}
\langle \ln |m_i|\rangle <-\alpha \ln 2   \llor  \langle \ln |m_i|\rangle < \langle \ln |R^\alpha_i|\rangle 
\Rightarrow \exists\,\text{1 MZM}\,.
\end{equation}
Let us discuss the effects of a small disorder on the topological phases in this case. 

\begin{figure}
[t!]
\centering
\includegraphics[width=0.49\columnwidth]{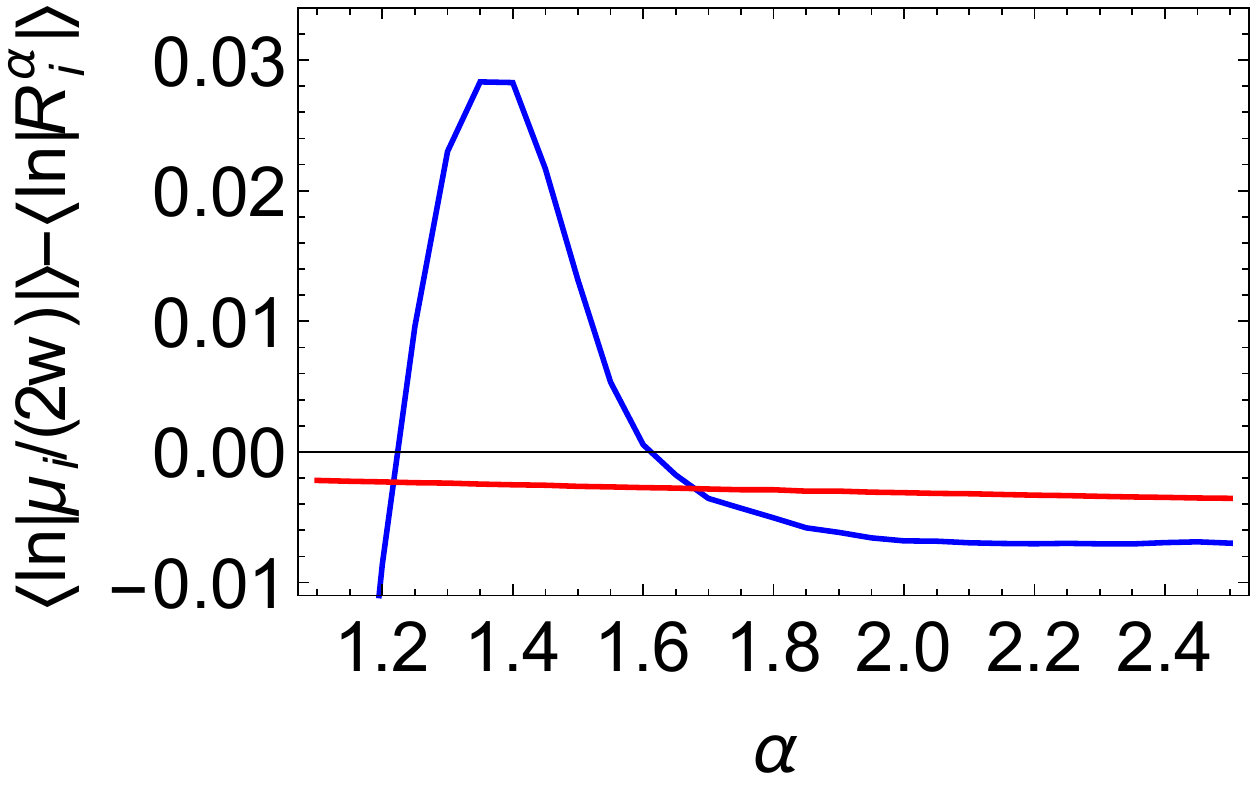}\hspace{0.1cm}\includegraphics[width=0.49\columnwidth]{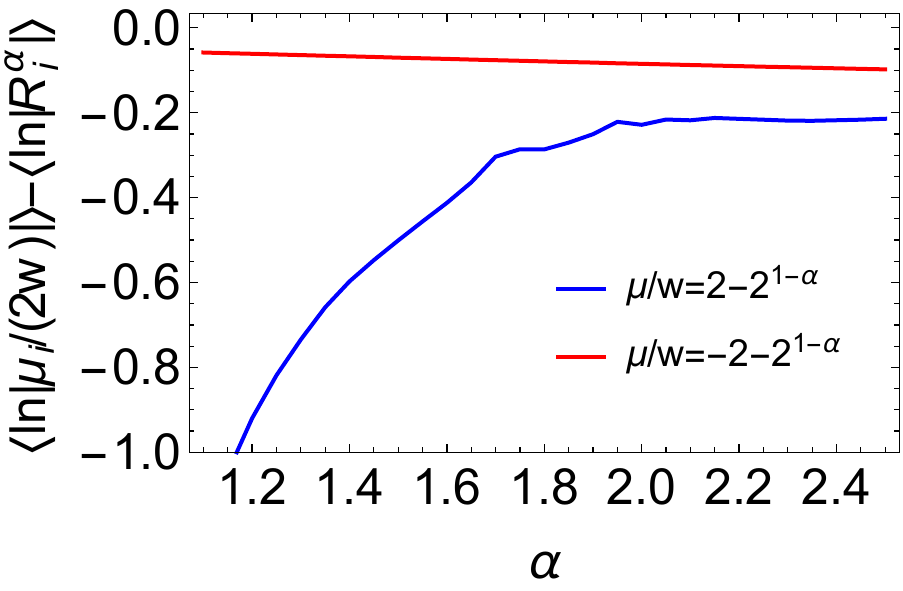}
\caption{$\langle\ln|m_i|\rangle -\langle \ln |R^\alpha_i|\rangle$,  calculated for $n=10^6$, as a function of $\alpha>1$, 
for a bimodal distributed disorder with strength $\epsilon=0.1$ in the left panel and $\epsilon=0.5$ in the right panel. The topological phase with 1 MZM per edge is for negative values, $\langle \ln|m_i| \rangle - \langle \ln|R_i| \rangle<0$.
}
\label{fig:topo-vs-alpha}
\end{figure}

For $\alpha>1$, instead, we can have 0 or 1 MZM per edge and we can consider only Eq.~\eqref{eq. c1 alpha}. In the absence of disorder, namely for $m_i=m$,  we get $\alpha \ln 2 + \langle \ln |m_i|\rangle > 0$, and from the condition  $\langle \ln |m_i|\rangle - \langle \ln |R^\alpha_i|\rangle <0 $ we obtain that there is 1 MZM for $(2^{-\alpha}-1)<m< (2^{-\alpha}+1)$. For simplicity, let us consider a bimodal distribution, i.e., $m_i=m+\epsilon$ and $m_i=m-\epsilon$ with same probability and focus on the points $m=\pm1 + 2^{-\alpha}$. As shown in Fig.~\ref{fig:topo-vs-alpha}, we get that the topological phase with 1 MZM can be suppressed, only at the boundary $m\approx -1 + 2^{-\alpha}$ (i.e. at $\mu/w\approx 2-2^{1-\alpha}$), for $\epsilon$ small enough, while it widens otherwise.

\begin{figure}
[t!]
\centering
\includegraphics[width=0.49\columnwidth]{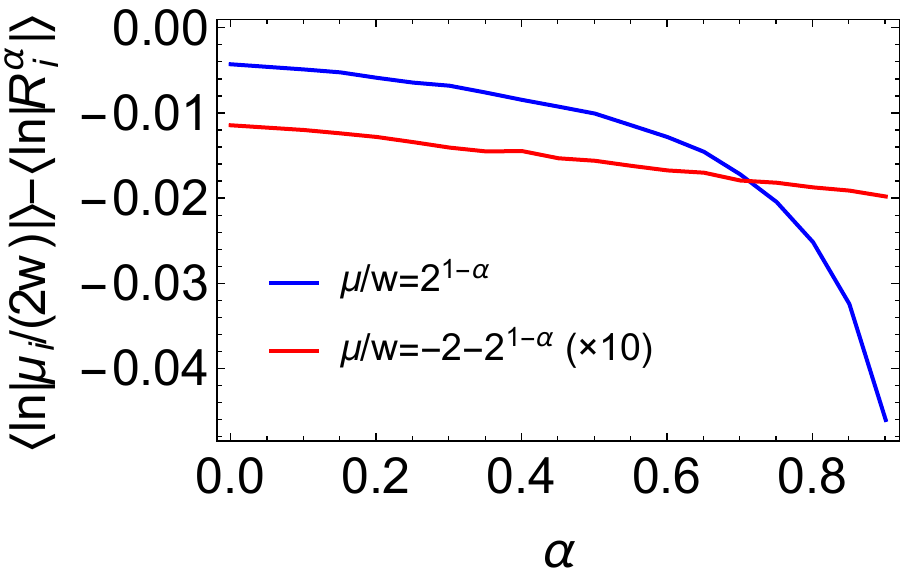}\hspace{0.1cm}\includegraphics[width=0.49\columnwidth]{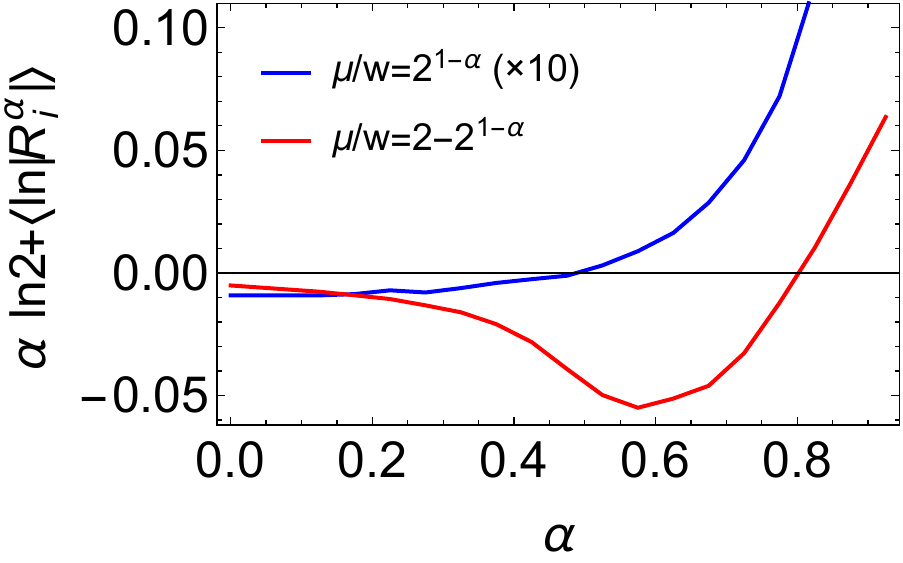}
\caption{$\langle\ln|m_i|\rangle -\langle \ln |R^\alpha_i|\rangle$ (left panel) and $\alpha \ln2 + \langle \ln |R^\alpha_i|\rangle$ (right panel) calculated for $n=10^6$, as functions of $\alpha<1$, for a bimodal distributed disorder with strength $\epsilon=0.1$. 
The topological phase with 1 MZM per edge is defined for $\langle \ln|m_i| \rangle - \langle \ln|R^\alpha_i| \rangle<0$ while the phase with 2 MZMs for $\alpha \ln2 + \langle \ln |R^\alpha_i|\rangle<0$.}
\label{fig:topo-vs-alpha-2}
\end{figure}

For $\alpha<1$, we can have 0, 1 or 2 MZMs. In the absence of disorder, i.e.,  for $m_i=m$, there is 1 MZM for $-2^{-\alpha}<m< (1+2^{-\alpha})$ and 2 MZMs if  $-2^{-\alpha}<m< (2^{-\alpha}-1)$.  Let us focus on the phase with 1 MZM  
in the presence of disorder, defined by the condition given in Eq.~\eqref{eq. c1 alpha}. For small $\epsilon$,  $\langle\ln|m_i|\rangle -\langle \ln |R_i|\rangle$ is negative at both boundaries, $m=-2^{-\alpha}$ and $m= 1 + 2^{-\alpha}$, that means that the topological phase widens (see Fig.~\ref{fig:topo-vs-alpha-2}).
On the contrary, concerning the phase with 2 MZMs, defined by the condition in Eq.~\eqref{eq. c2 alpha}, for small $\epsilon$, $\alpha \ln 2 + \langle \ln |m_i|\rangle<0$ at both boundaries $m=-2^{-\alpha}$ and $m= 2^{-\alpha}-1$, while $\alpha \ln2 + \langle \ln |R_i|\rangle$ is positive only for large $\alpha$, so that in this case the topological phase shrinks (see Fig.~\ref{fig:topo-vs-alpha-2}).

\section{Many-neighbor couplings ($r>2$)}

Similarly to what we have done for $r=2$, let us consider, for simplicity, $\Delta_\ell=w_\ell$, starting first 
with the extreme case where $w_\ell = 0$ for any $\ell<r$ but $w_r\neq 0$. 
From Eq.~\eqref{eq. eig. 0} we get
\begin{equation}
\phi_{i+r}+\frac{\mu_i}{2w_r}\phi_i =0\,,
\end{equation}
from which we obtain $r$ equations for $i_0=1,2, \dots, r$ 
\begin{equation}
\phi_{r n+i_0}=\prod_{i=0}^{n-1} \left(-\frac{\mu_{r i+i_0}}{2 w_r}\right)\phi_{i_0}\,,
\end{equation}
which generalizes Eqs.~(\ref{phin1}) and (\ref{phin2}). 
We have now $r$ copies of the condition for getting one MZM
\begin{equation}\label{eq. con s}
\left\langle \ln \left| \frac{\mu_{i}}{2 w_r}\right|\right\rangle_{i_0}  < 0\,,
\end{equation}
where we define the averages $\langle x_i \rangle_{i_0} = 
r/n \sum_{i} x_{ri + i_0}$, 
valid for large $n$.
In detail, $i=1,\cdots,r$ so that we have $r$ conditions.
If locally the distribution probability of the disorder is the same at any site, the $r$ conditions in Eq.~\eqref{eq. con s} (with $i_0=1,\dots,r$) are equal, and we can get only zero or $r$ MZMs per edge. For general couplings, instead, we can have an intermediate number of MZMs. 

Let us now consider non-vanishing $w_{\ell}$ focusing our attention, for simplicity, always to the case $w_\ell=\Delta_\ell$.\\ 
For $r=3$, for instance, 
in order to determinate the eigenvalues $\lambda_n$ of $\tilde A_n$, we have to solve the following equation
\begin{equation}\label{eq. lambda 1 r3}
\lambda_n^3 - \lambda_n^2 T_n + \lambda_n T'_n-D_n = 0\,,
\end{equation}
where, to simplify notation, we define $T_n=\text{Tr}(\tilde A_n)$, $T'_n=((\text{Tr}(\tilde A_n))^2-\text{Tr}(\tilde A_n^2))/2$ and $D_n=\det(\tilde A_n)$. From Eq.~(\ref{eq. lambda 1 r3}) we derive the following conditions
\begin{eqnarray}
 \label{eq. c3 r3} &&D_n\to 0 \lland T'_n\to 0 \lland T_n \to 0 \Rightarrow \exists\,\text{3 MZMs}\,,\\
 \label{eq. c2 r3} \nonumber &&(D_n \to 0 \lland  T'_n\to 0) \;\textrm{or}\;
 \left(\frac{D_n}{T_n}\to 0 \lland \frac{T'_n}{T_n}\to 0 \right) \\
\label{eq. c2 r3} &&{\phantom{-----------------}}\Rightarrow \exists\,\text{2 MZMs}\,,\\
 \label{eq. c1 r3} &&D_n\to 0 \llor \frac{D_n}{T_n}\to 0 \llor \frac{D_n}{T'_n}\to 0 \;\Rightarrow \exists\,\text{1 MZM} \,.
\end{eqnarray}
As discussed in Appendix~\ref{app. r3}, we expect that these conditions can be also expressed in terms of random continued fractions analogously to the next-nearest-neighbor case.%

Equations~\eqref{eq. c3 r3}-\eqref{eq. c1 r3} can be generalized for an arbitrary $r$, as shown in Appendix~\ref{app. any r}.
In particular, we note that the condition in Eq.~\eqref{eq. c3 r3} can be easily generalized to
\begin{equation}
\det(\tilde A_n)\to 0 \lland \text{Tr}(\tilde A_n^k)\to 0 , \forall k\in I_{1,r-1} \Rightarrow \exists\,\text{r MZMs}\,,
\end{equation}
where $I_{i,j}=\{i,i+1,\cdots,j\}$.
Furthermore, for an arbitrary range $r$, $\det(\tilde A_n)\to 0$ is a sufficient (but not necessary) condition for getting 1 MZM per edge, and since $|\det(\tilde A_n)|\sim |w_1/w_r|^n e^{n \langle \ln|m_i| \rangle }$, we get that the condition $\langle \ln|m_i| \rangle<-\ln|w_1/w_r|$ implies that there is 1 MZM. As a result, if $|w_1|\geq |w_r|$, the topological phase with 1 MZM for $r>1$ cannot be smaller than that for $r=1$, for the same disorder and coupling $w_1$. 

\begin{figure}
[t!]
\centering
\includegraphics[width=0.495\columnwidth]{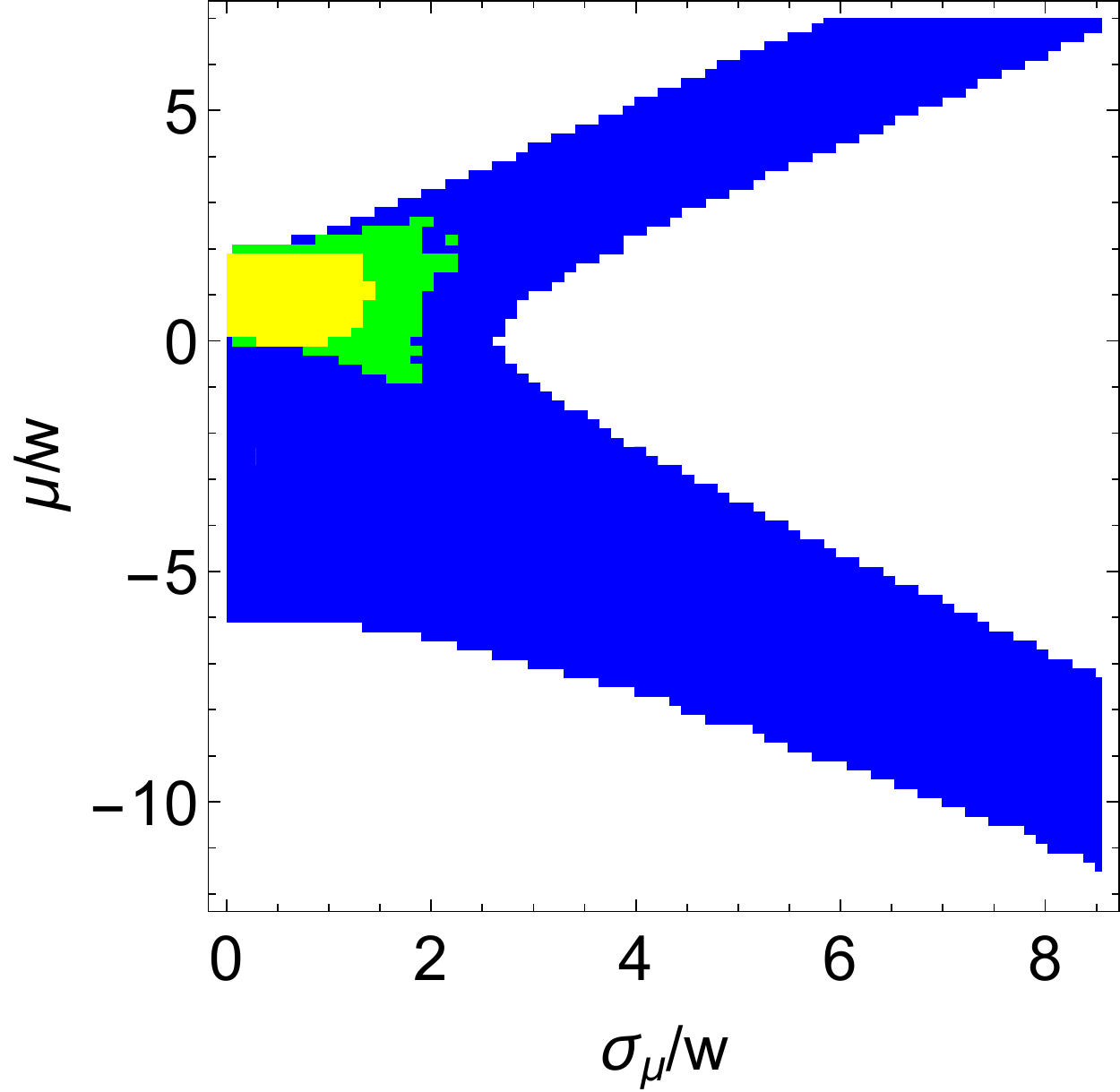} \includegraphics[width=0.495\columnwidth]{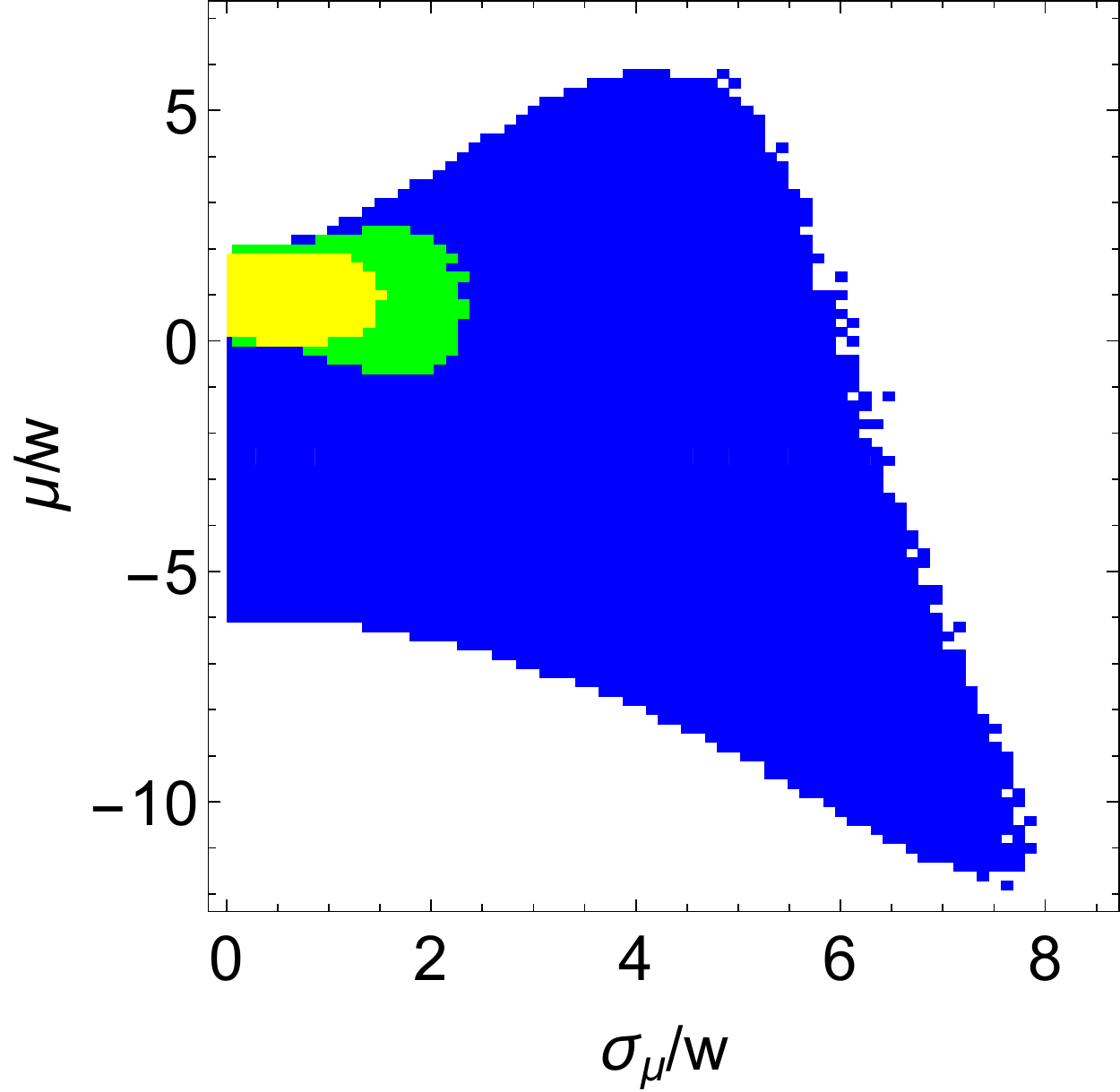}
\includegraphics[width=0.495\columnwidth]{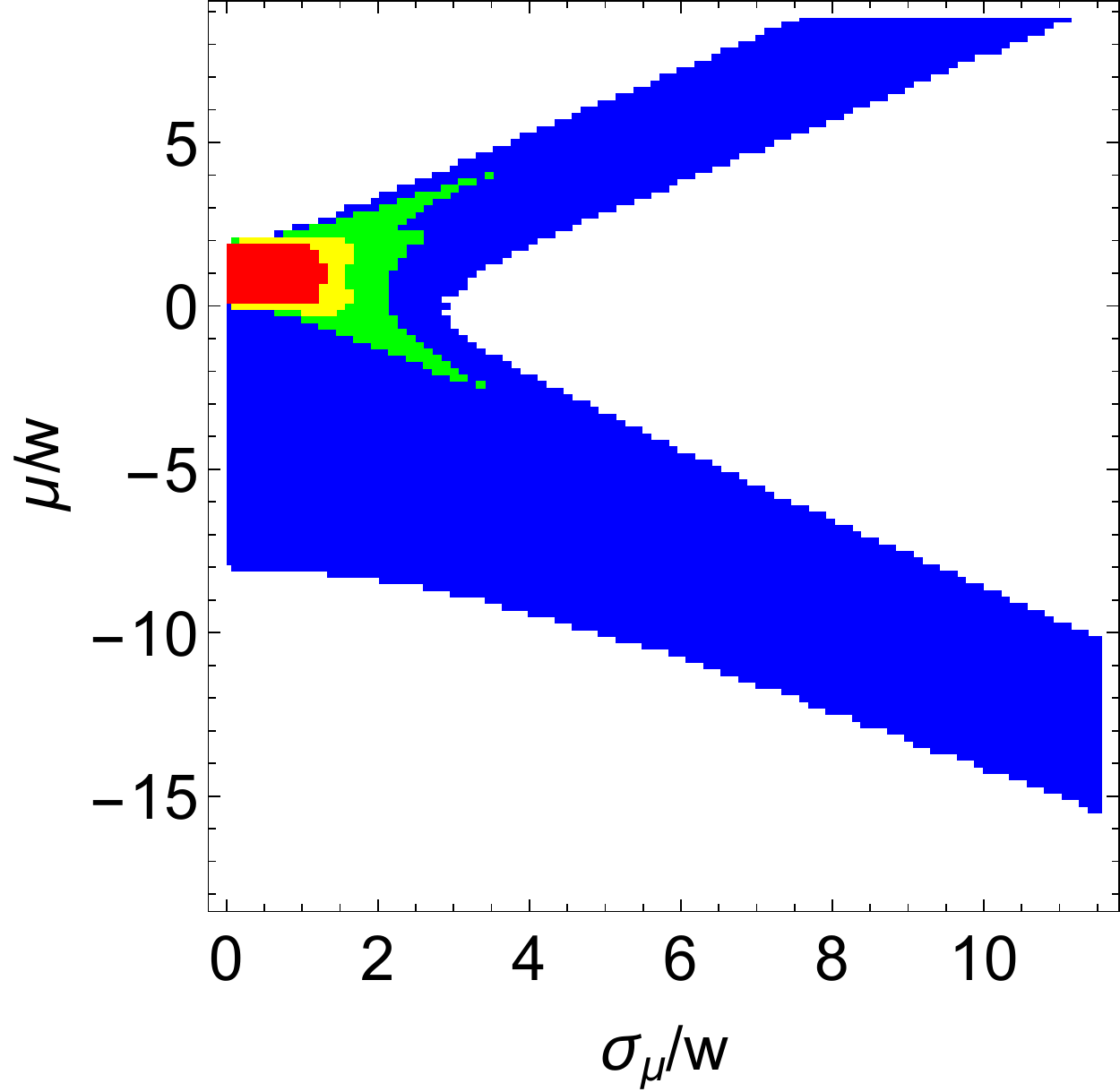} \includegraphics[width=0.495\columnwidth]{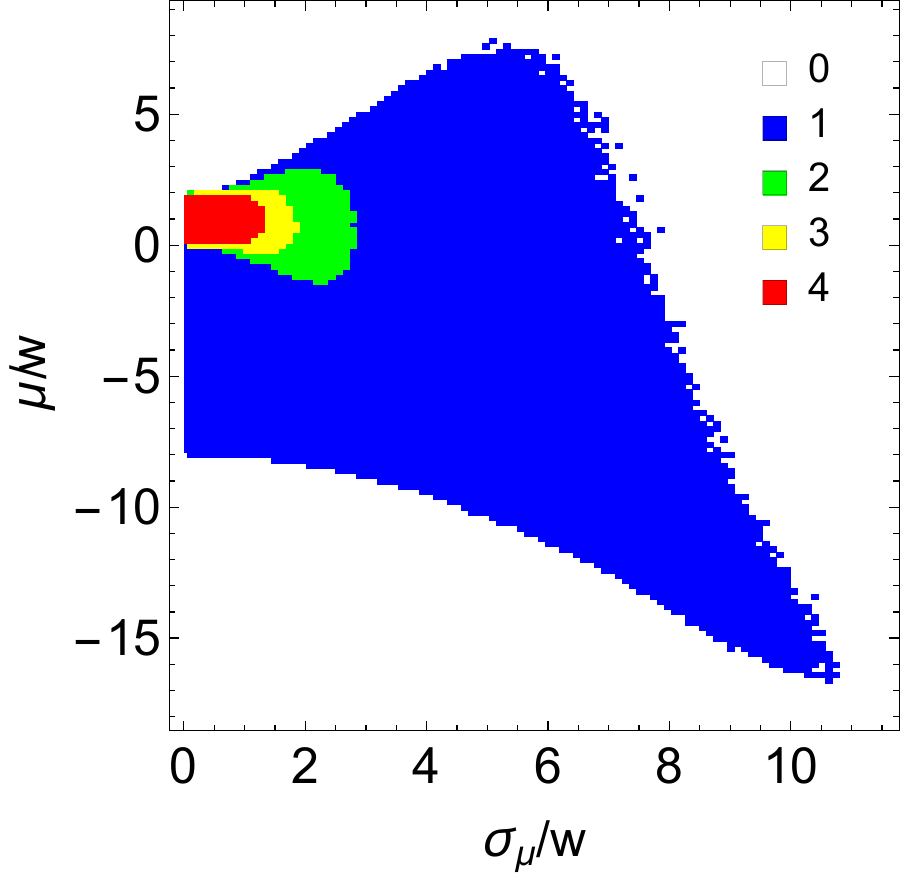}
\caption{Topological phase diagrams $\sigma_\mu$-$\mu$, for $r=3$ (upper panels) and $r=4$ (lower panels), for $w=w_1=w_2=\Delta_1=\Delta_2$, with $L=200$. The diagrams are characterized by $0$ (white), $1$ (blue), $2$ (green), $3$ (yellow), and $4$ (red) MZMs per edge, for a bimodal  distribution (left panels) and a uniform disorder distribution (right panels).}
\label{fig:topo r>2}
\end{figure}

Finally, for $w_\ell=w$, without disorder there are 0, 1 or $r$ MZMs (see Ref.~\cite{alecce17}). In detail, there are $r$ MZMs if $m\in (-1,0)$ and there is 1 MZM if $m \in (0,r)$. In the presence of the disorder, instead, we can have many phases, with a number of MZMs which goes from 0 to $r$, as shown in Fig.~\ref{fig:topo r>2}.
Moreover, we note that a large disorder significantly increases the topological regions.

\section{Infinite number of neighbors}
To investigate the limit $r\to \infty$ corresponding to an infinite number of neighbors, we consider the following algebraically decaying parameters $w_\ell=w/\ell^\alpha$ and $\Delta_\ell = \Delta/\ell^\beta$. In the absence of disorder, namely for a homogeneous chemical potential $\mu_i=\mu$, and 
for $\alpha$ and $\beta$ greater than 1, in the thermodynamic limit, we can have 0 or 1 MZM per edge, if the Majorana number $\mathcal M = \text{sgn}((\mu+g(0) w)(\mu+g(\pi)w))$ is 1 or -1, where $g(k)= 2 \textrm{Re}[Li_\alpha(e^{ik})]$, with $Li_\alpha(x)$ the polylogarithm. In the presence of disorder, we expect also that, for exponents larger than one, there are 0 or 1 MZM per edge, as for finite $r$. %

We consider an uncorrelated box distribution, namely where the random variables are uniformly distributed, $m_i=m+ \epsilon x_i$ with $x_i$ uniform in $[-1,1]$.
For an open chain, described by the Hamiltonian $H=\frac{1}{4} \sum_{l,m} c_{l} {\cal H}_{lm}c_m$, as that in Eq.~\eqref{eq. Hami}, we can calculate the following topological invariant 
\be
\label{eq.loring}
W=\text{Sig}(X \Gamma + {\cal H})
\ee
as defined in Ref.~\cite{loring15}. Specifically, the signature $\text{Sig}(M)$ of a matrix $M$ is defined as the number of positive eigenvalues of $M$ minus the number of negative ones. We define the position operator  $X=\text{diag}(X_j)\otimes I_2$, where the positions $X_j$ are normalized such that $-1/2\leq X_j\leq 1/2$, and the grading operator $\Gamma=I_L\otimes \tau_3$, where $\tau_i$ with $i=1,2,3$ are the Pauli matrices and $I_n$ the $n\times n$ identity matrix.%
\begin{figure}
\includegraphics[width=0.475\columnwidth]{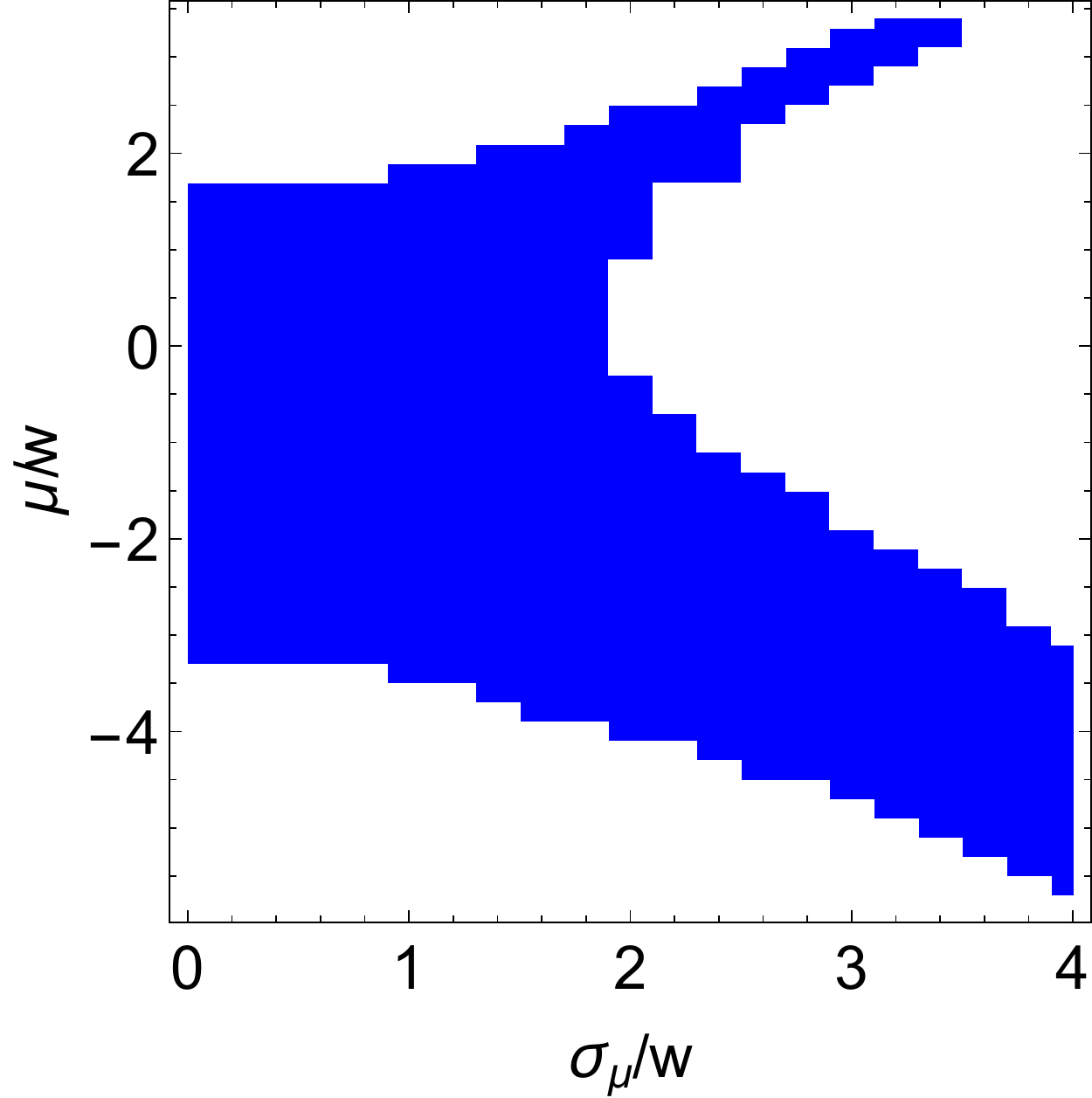}
\includegraphics[width=0.485\columnwidth]{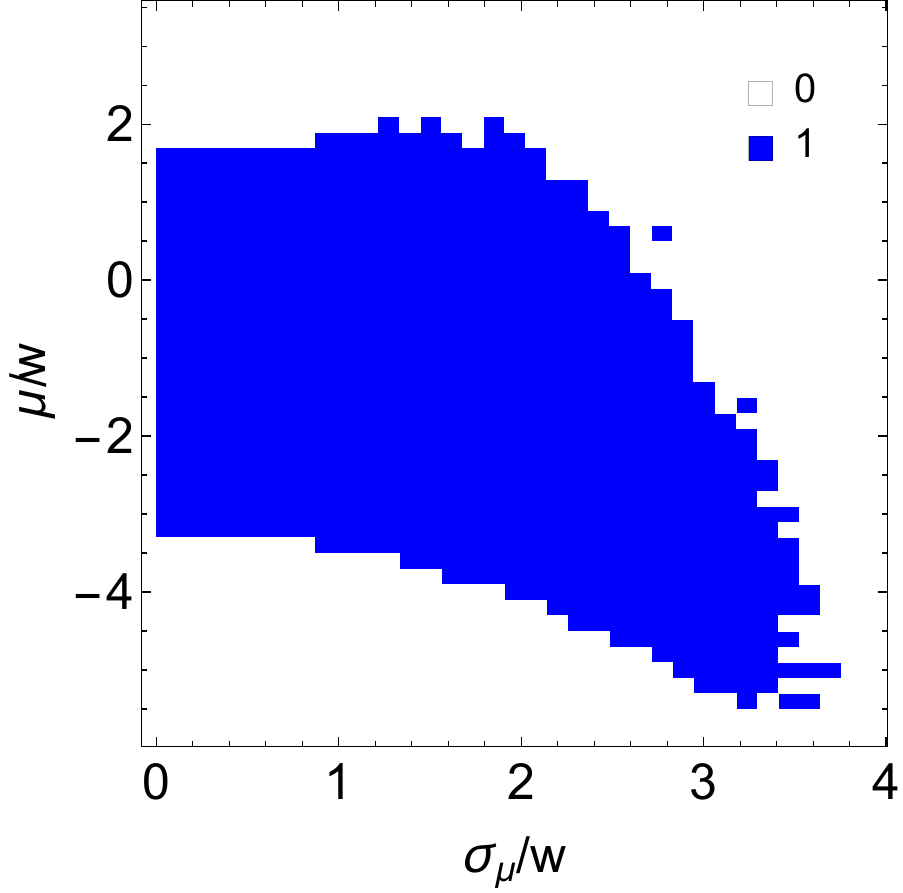}
\caption{Topological phase diagram $\sigma_\mu$-$\mu$ for long-range couplings $w_\ell=\Delta_\ell=w/\ell^\alpha$ with $\alpha=2$, obtained considering a chain with size $L=200$ and averaging over $100$ disorder realizations, with bimodal distribution (left), with $m_i=m\pm \epsilon$, namely $\mu_i=\mu\pm \sigma_\mu$, and uniform box distribution (right), $m_i\in[m-\epsilon,m+\epsilon]$, namely $\mu_i \in [\mu-\sqrt{3}\sigma_\mu,\mu+\sqrt{3}\sigma_\mu]$, characterized by 0 (white region) and 1 (blue region) MZM. }
\label{fig:loring1}
\end{figure}
\begin{figure}
\includegraphics[width=0.49\columnwidth]{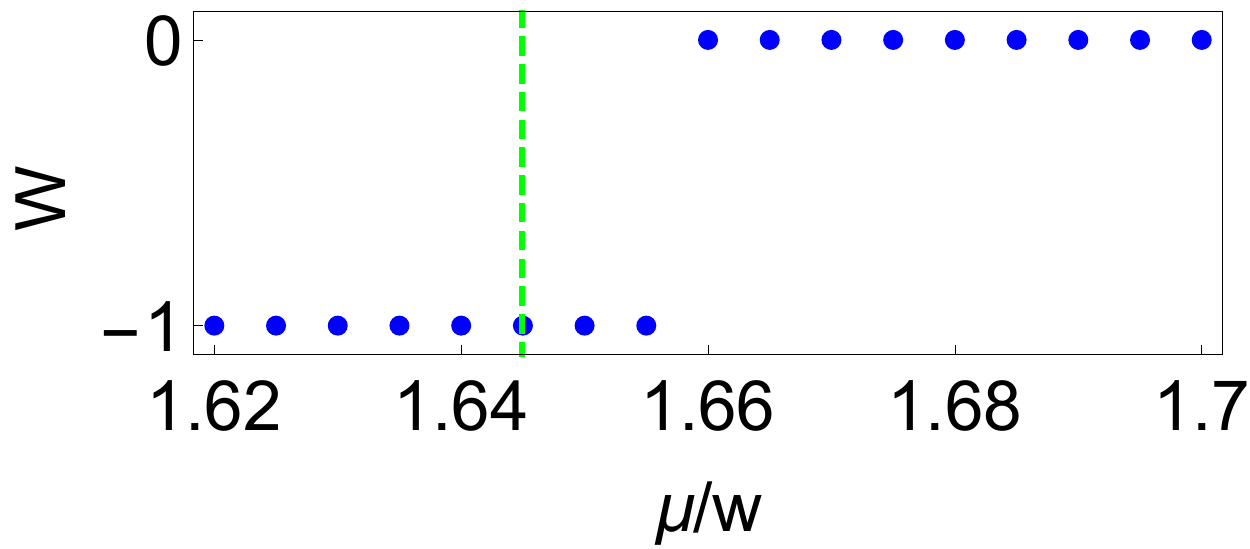} 
\includegraphics[width=0.49\columnwidth]{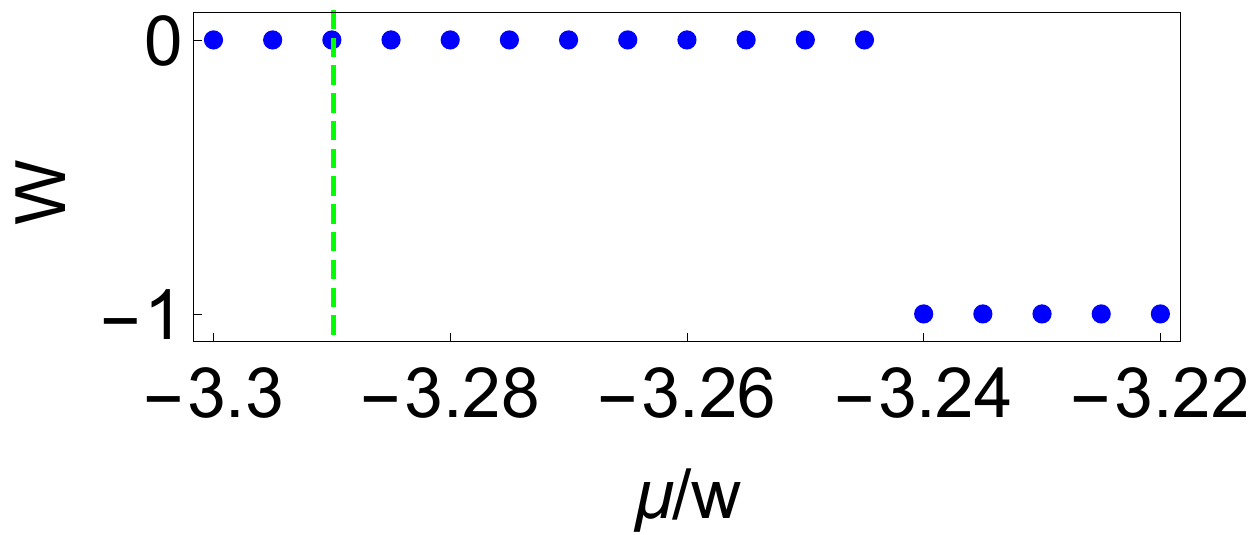}
\caption{
Topological invariant $W$, Eq.~(\ref{eq.loring}), as a function of $\mu$, for $L=200$, and $w_\ell=\Delta_\ell=w/\ell^\alpha$ with $\alpha=2$ at fixed small disorder $\epsilon=0.2$ (fixed $\sigma_\mu/w=2\epsilon/\sqrt{3}$), close to the two boundaries for a clean system. $W=0$ corresponds to a trivial phase, $W=-1$ to a topological phase. The green dashed lines correspond to the critical values in the absence of disorder.}
\label{fig:loring2}
\end{figure}

We find that either a discrete bimodal distribution and a uniformly distribution of the disorder, on a chain with long range coupling $w_\ell=\Delta_\ell=w/\ell^\alpha$, for $\alpha>1$, induce qualitatively similar behaviors as those found for a finite number of neighbors, supporting and widening the topological phases, see Fig.~\ref{fig:loring1}. As shown in the two panels of Fig.~\ref{fig:loring1}, for both distributions, the topological phases are not symmetric, as for the cases with finite $r>1$, for relatively small $\alpha$. They become symmetric around $\mu=0$ for $\alpha\gg1$, recovering the short-range results reported in Fig.~\ref{fig:topo r=1}. 

Let us now discuss in more detail the small disorder regime. 
As shown in Fig.~\ref{fig:loring2}, where $W$ is reported varying $\mu$, always for $w_\ell=\Delta_\ell$ and for very small disorder, the topological phase widens for $\mu\approx - g(\pi) w>0$ (top panel) and slightly shrinks for $\mu\approx - g(0) w<0$ (bottom panel) for $\alpha$ not too large. 
However, since the limit $\alpha\to\infty$ is equivalent to the case $r=1$, we expect that, for $\alpha$ large enough, the topological region widens again at both boundaries.  

In general, this behavior can be explained by considering the effective Hamiltonian
\begin{equation}
H_e = H_0 + \Sigma\,,
\end{equation}
where $\Sigma$ is the self-energy defined such that
\begin{equation}
\langle G\rangle = G_e\,,
\end{equation}
where we have defined the Green's functions $G=(E-H)^{-1}$ and $G_e=(E-H_e)^{-1}$, and where the average is done over disorder. $H_0$ is the clean Hamiltonian as in Eq.~(\ref{eq. Hami}) with homogeneous $\mu_i=\mu$, %
so that $H=H_0 + H_1$, where $H_1$ is the disorder term such that $\langle H_1 \rangle = 0$. To determinate the self-energy, we define $H'_1 = H_1-\Sigma$, then the Green's function reads
\begin{equation}
G=\frac{1}{E-H_e-H'_1}\,,
\end{equation}
and by considering $H'_1$ as a small perturbation, we get at the second order
\begin{equation}
G \approx G_e (1+H'_1 G_e +H'_1 G_e H'_1 G_e)\,.
\end{equation}
By taking the average of this equation, we get
\begin{equation}
\langle H'_1\rangle + \langle H'_1 G_e H'_1\rangle \approx 0\,,
\end{equation}
and since $\langle H'_1 \rangle = - \Sigma $ we have
\begin{equation}
\Sigma \approx \langle H'_1 \frac{1}{E-H_0-\Sigma}H'_1 \rangle\,.
\end{equation}
Since we are interested on small disorder, we consider $\Sigma$ as a second order correction, thus we get
\begin{equation}\label{eq. s e}
\Sigma \approx \langle H_1 G_0 H_1 \rangle\,,
\end{equation}
where $G_0=(E-H_0)^{-1}$.
For our model, the self-energy $\Sigma$, at $E=0$ in the bulk, reads (see Appendix~\ref{app. s e})
\begin{eqnarray}
\nonumber\Sigma &\approx& -\frac{i}{2} \sum_j \delta\mu \,c_{2j-1}c_{2j} + \frac{i}{2} \sum_{\ell,j}\big[(\delta w_\ell +\delta \Delta_\ell)c_{2j} c_{2j+2\ell-1}\\
\label{eq. se s} && -(\delta w_\ell -\delta \Delta_\ell)c_{2j-1} c_{2j+2\ell}\big].
\end{eqnarray}
Adding $\Sigma$ to $H_0$ we obtain the effective Hamiltonian $H_e$ whose parameters are 
an effective chemical potential, $\mu+\delta\mu$, where%
\begin{equation}
\delta \mu = - \frac{\sigma_\mu^2}{\pi} \int_0^\pi dk \frac{\mu + w g(k)}{(\mu+w g(k))^2+(\Delta f(k))^2}\,,
\label{dmu}
\end{equation}
an effective long-range hopping, $w_\ell+\delta w_\ell$, where%
\begin{equation}
\delta w_\ell = - \frac{\sigma_\mu^2}{\pi} \int_0^\pi dk \cos(k\ell)\frac{\mu + w g(k)}{(\mu+w g(k))^2+(\Delta f(k))^2}\,,
\label{dwl}
\end{equation}
and an effective pairing, $\Delta_\ell+\delta \Delta_\ell$, where%
\begin{equation}
\delta \Delta_\ell = - \frac{\sigma_\mu^2}{\pi} \int_0^\pi dk \sin(k\ell)\frac{\mu + w g(k)}{(\mu+w g(k))^2+(\Delta f(k))^2}\,,
\label{dDl}
\end{equation}
where $\mu$ and $\sigma_\mu^2$ are the mean value and the variance of the disorder in the chemical potential, $g(k)= 2 \textrm{Re} [Li_\alpha(e^{ik})]$ and $f(k)= 2 \textrm{Im}[Li_\beta(e^{ik})]$.
In order to characterize the topological phases, we can use the Majorana number of the effective Hamiltonian $H_e$. 

Let us consider $w_\ell=\Delta_\ell$. For $\mu = - w g(0)<0$, we get $\delta \mu >0$, and $\delta w_\ell \approx -c'\delta \mu \ell^{-\alpha'}$ where $\alpha'$ increases with $\alpha$, and where $c'>0$. We can, then, calculate the Majorana number
\begin{equation}
\mathcal M \approx  \text{sgn}[(\delta \mu - c' \delta \mu g'(0))( w g(\pi)- w g(0))]\,,
\end{equation}
where $g'(k)=\sum_\ell \cos(k\ell) \ell^{-\alpha'}= 2 \textrm{Re} [Li_{\alpha'}(e^{ik})]$. We have that $w g(\pi)- w g(0)<0$ and $1-c'g'(0)<0$ for any $\alpha'$, if $c'>1/2$, or for $\alpha'$ not too large, if $c'<1/2$. We find that $c'$ is small, so that $\mathcal M >0$ (trivial phase) for $\alpha$ not too large.
On the other hand, for $\mu = - w g(\pi)>0$, we get $\delta \mu <0$, and $\delta w_\ell \approx c''\delta \mu (-1)^{\ell}\ell^{-\alpha''}$ where $c''>0$. Then, we get
\begin{equation}
\mathcal M \approx \text{sgn}[(\delta \mu + c''\delta \mu g''(\pi))( w g(0)-w g(\pi))]\,,
\end{equation}
where $g''(k)= \sum_\ell \cos(k\ell) (-1)^\ell \ell^{-\alpha''}$. We have that $g''(\pi)>0$ for any $\alpha''$, so that $\mathcal M <0$ (topological phase) for any $\alpha$. 

\begin{figure}[t!]
\centering
\includegraphics[width=0.48\columnwidth]{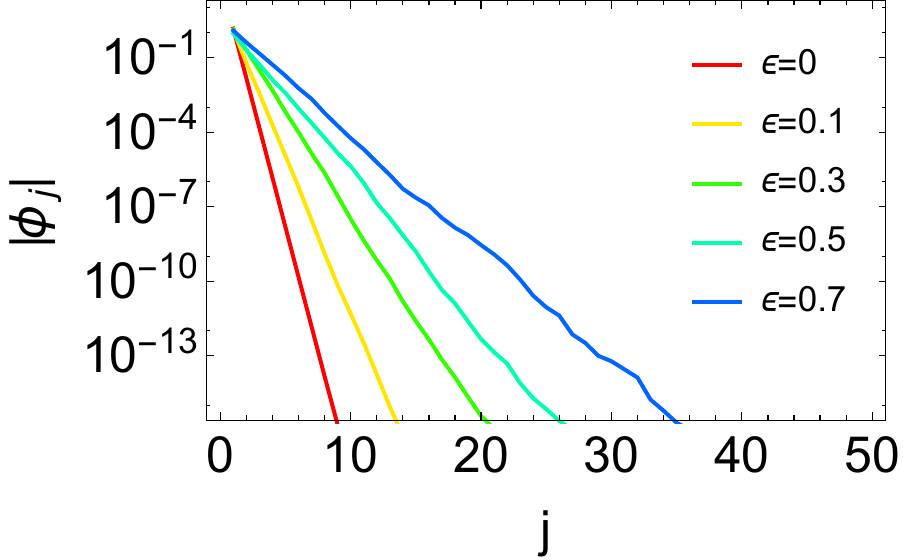}\hspace{0.1cm}\includegraphics[width=0.48\columnwidth]{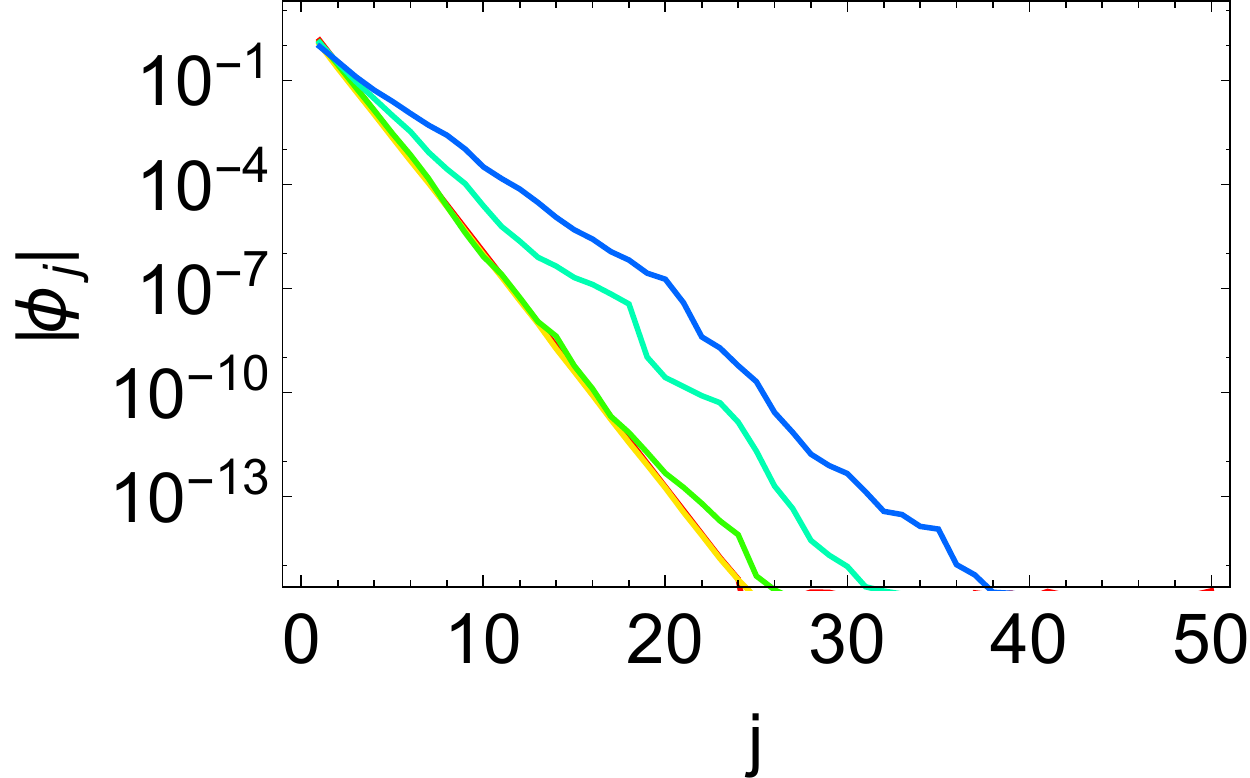}
\includegraphics[width=0.48\columnwidth]{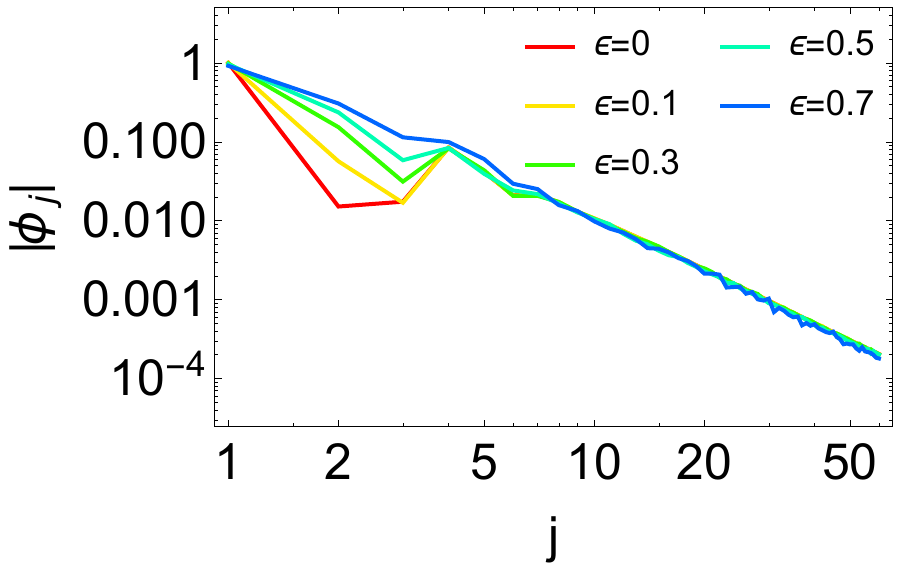}\hspace{0.1cm}\includegraphics[width=0.48\columnwidth]{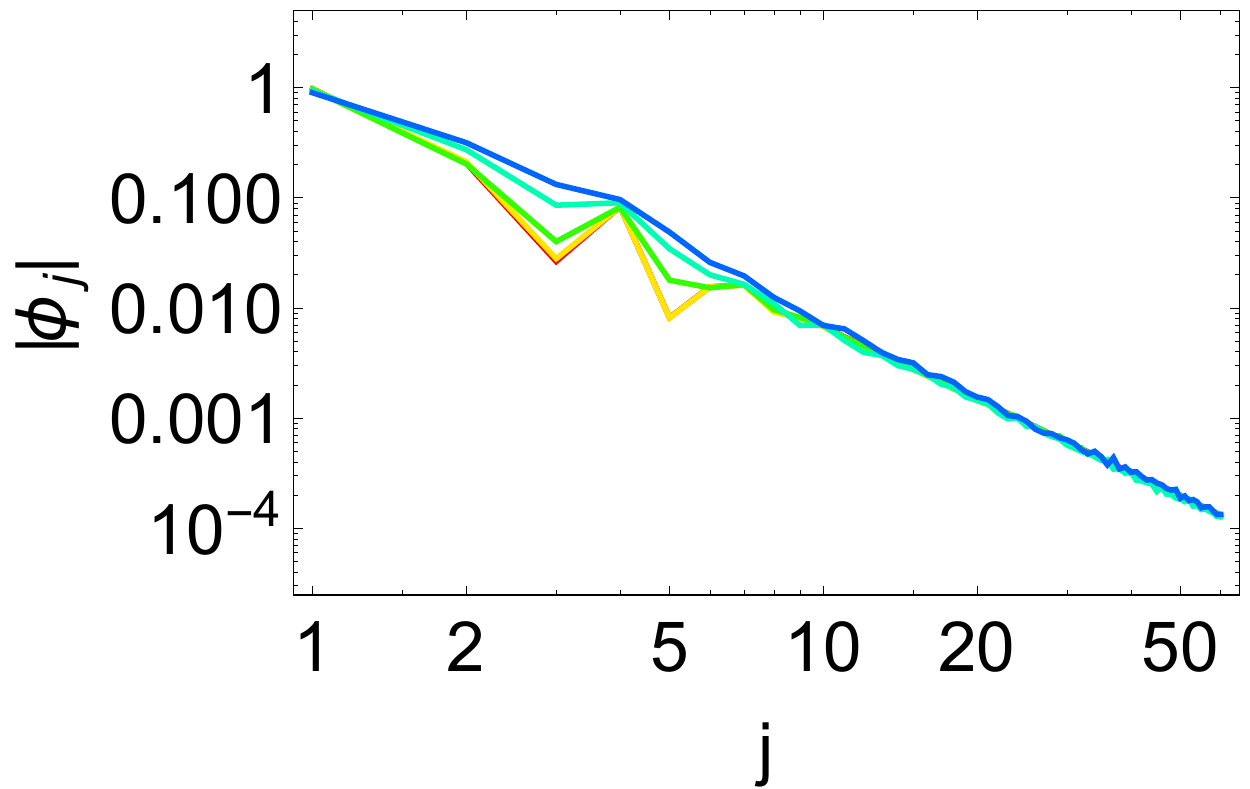}
\caption{(Top) Wave-function $\phi_j$ of a MZM, as a function of the distance $j$ from the edge, for $L=200$, $w_\ell=\Delta_\ell=w/\ell^\alpha$ with $\alpha=2$, for different values of the disorder strength $\epsilon$, at $\mu/(2w)=0.01$ (left panel) and $\mu/(2w)=0.2$ (right panel). We average over 100 disorder realizations. The plots are in linear-log scale, showing exponential decays. 
(Bottom) 
Wave-function $\phi_j$ of a MZM, as a function of the distance $j$ from the edge, for $L=200$, with $w_\ell\neq \Delta_\ell$, in particular $w_\ell=w/\ell^\alpha$ and $\Delta=w/\ell^\beta$ with $\alpha=4$ and $\beta=2$. We consider different values of the disorder strength $\epsilon$, at $\mu/(2w)=0.01$ (left panel) and $\mu/(2w)=0.2$ (right panel). The plots are in log-log scale, showing algebraic decays. 
}
\label{fig:loc psi}
\end{figure}

In the end, we note that for $w_l\neq\Delta_l$ and $w_l=\Delta_l$ the MZMs wave-functions remain algebraically and exponentially localized, respectively, also in the presence of disorder (see Fig.~\ref{fig:loc psi}).
This can be explained by using the method of Ref.~\cite{francica22}. By considering the self-energy for small disorder, we get the wave function of a MZM
\begin{equation}
\phi_j \sim \textrm{Im} \oint_{|z|=1} \frac{z^{j-1}dz}{\mu+\delta \mu + 
w\tilde g(z)+\Delta \tilde f(z)
+ Y(z)}\,
\end{equation}
where $\tilde g(z)=(Li_\alpha(z)+Li_\alpha(1/z))$, $\tilde f(z)=(Li_\beta(z)-Li_\beta(1/z))$ and where $Y(z)$ is defined by performing an analytical continuation in the complex plane such that 
\be
Y(e^{ik}) = \sum_\ell \big(\delta w_\ell\cos(k\ell)+i \delta \Delta_\ell \sin(k\ell)\big).
\ee 
$Y(z)$ has no branch cut for $|z|<1$. %
Then, for $\Delta_\ell\neq w_\ell$ we get a branch cut coming from the polylogarithms. As a result, we get an algebraically decay of the wavefunction,  
while for $\Delta_\ell = w_\ell$ there is no branch cut and we get a purely exponential decay, as in the absence of disorder \cite{francica22,jager20}. We note that since the corrections $\delta w_\ell $ and $\delta \Delta_\ell $, and therefore the effective couplings, are different, also for $\Delta_\ell=w_\ell$, 
one might expect that the MZMs become algebraically localized in the presence of disorder. However, this case does not occur because of the analytic properties of the function $Y(z)$. In particular, for $\Delta_\ell=w_\ell$, the MZMs stay exponentially localized with a localization length which increases upon increasing the strength of the disorder (see Fig. \ref{fig:loc psi}, top panels). In the other case, for $\Delta_\ell\neq w_\ell$ the MZMs remain algebraically localized  (see Fig. \ref{fig:loc psi}, bottom panels) as in the absence of disorder.

\section{Conclusions}
We carried out a complete study of the combined effects related to the interplay of disorder and range of the couplings on the topological phase diagrams for a one-dimensional topological superconductor. %
We considered both a finite and an infinite number of neighbors involved in the couplings. 

{\color{black} Our work is in perfect agreement with what is known in the previous literature, namely that disorder can stabilize the topological order (see, for instance, Ref. \cite{Adagideli14} for one-dimensional models and Ref. \cite{Haim19} for the multichannel models). In particular, we generalize what has been found in Ref. \cite{gergs16} for a one-dimensional topological superconductor described by a short-range Kitaev chain in the presence of disorder, by considering longer finite-range and infinite-range couplings, with two different kinds of random distributions. We show that both the coupling range and the disorder can promote the topological order.}

For finite-range couplings, we resorted to a transfer matrix approach which allowed us to derive the conditions for the existence of Majorana zero modes, and to show that either the range and the on-site disorder can greatly enhance the topological phases characterized by the appearance of one or many Majorana modes localized at the edges. 
Moreover, we discussed the role of correlated disorder which can further widen the topological regions. 
We always performed a comparing analysis considering a discrete and a continuous disorder distribution. In the former case the topological regions extend to regimes of infinitely strong disorder. 
Finally, we show that, for long-range couplings, the spatial decay of the edge modes remains algebraic or exponential for equal hopping and pairing, even in the presence of disorder. 

In conclusion, our results show that the combined effects of disorder and range of the couplings 
can generate non-trivial behaviors, increasing the number of phases and greatly widening their extension.

\subsection*{Acknowledgements}
The authors acknowledge financial support from the project BIRD 2021 "Correlations, dynamics and topology in long-range quantum systems" of the Department of Physics and Astronomy, University of Padova, and 
and from the European Union-Next Generation EU within the "National Center for HPC, Big Data and Quantum Computing" (Project No. CN00000013, CN1 Spoke 10 - Quantum Computing).
\appendix

\section{Trace of $\tilde A_n$ for $r=2$}\label{app. trace}

Let us calculate $\tilde A_n$ for $r=2$ considering the case with $\Delta_\ell=w_\ell$, with $w_1=w_2$, and $n$ even. By using the multiplication law
\begin{equation}
A_2 A_1 = \text{diag}(m_2,m_1)-A_1
\end{equation}
(which, however, can be easily generalized for $w_1 \neq w_2$ to $A_2 A_1 = \text{diag}(m_2,m_1)-t_1A_1\,$, 
with $m_i=-\mu_i/(2w_2)$ and $t_1=-w_1/w_2$), 
where $\text{diag}(m_2,m_1)$ is the diagonal matrix with elements $m_2$ and $m_1$, we get
\begin{equation}
\tilde A_n = d_{n,1} - A_{n}\, d_{n-1,1} - A_{n-1} \,d_{n-2,1}-\cdots - A_3 \,d_{2,1}  -A_1\,,
\end{equation}
where we define $d_{2,1} = \text{diag}(m_2,m_1)$, $d_{3,1}=\text{diag}(m_3,m_1)$, $d_{4,1}=\text{diag}(m_4,m_1)+\text{diag}(m_4,m_3) \,\text{diag}(m_2,m_1)$ and so on.  
In general terms, $d_{i,1}$ is the diagonal matrix obtained by summing all the products of the form $\text{diag}(m_i,m_{i_k})\,\text{diag}(m_{i_k-1},m_{i_{k-1}})\cdots \,\text{diag}(m_{i_1-1},m_{1}) $ where $i_j>i_{j-1}+1$ and $i_1>2$.
The trace is
\begin{eqnarray}
\nonumber\text{Tr}(\tilde A_n) &=& \text{Tr}(d_{n,1}) - \text{Tr}(A_{n} d_{n-1,1}) - \text{Tr}(A_{n-1} d_{n-2,1})-\cdots\\
 && - \text{Tr}(A_3 d_{2,1})  -\text{Tr}(A_1)\,.
\end{eqnarray}
Since $\text{Tr}(A_i \,\text{diag}(a,b)) = - a$, we get
\begin{equation}
\text{Tr}(\tilde A_n) = \text{Tr}(d_{n,1}) + d^{1}_{n-1,1} + d^{1}_{n-2,1}+\cdots + d^{1}_{2,1} +1\,,
\end{equation}
where $d^{1}$ is the element $(1,1)$ of the diagonal matrix $d$. In particular
\begin{equation}
d^{1}_{i,1} = \sum_l p^{u}_i(l)%
\end{equation}
where $p^{u}_i(l)$ %
are all the products of the upper endpoints of the non-crossing partitions of  $(m_i,m_{i-1},\cdots,m_1)$ with at least two elements, e.g., the non-crossing partitions of $(m_5,m_4,m_3,m_2,m_1)$ are $(m_5,m_4,m_3,m_2,m_1)$, $(m_5,m_4)\cup (m_3,m_2,m_1)$ and $(m_5,m_4,m_3)\cup (m_2,m_1)$, then $d^{1}_{5,1}=m_5 + m_5 m_3 + m_5 m_2$. We get the relation
\begin{equation}
d^1_{i,1}=m_i(1+d^1_{i-2,1}+d^1_{i-3,1}+\cdots +d^1_{2,1})
\end{equation}
By defining $p^{\ell}_i(l)$ %
as all the products of the lower endpoints, we get
\begin{equation}
d^{2}_{i,1} = \sum_l p^{\ell}_i(l)%
\end{equation}
Since 
$\text{Tr}(d_{i,1}) = d^{1}_{i,1}+d^{2}_{i,1}$, 
the trace of $\tilde A_n$ is, then, given by 
\begin{equation}
\text{Tr}(\tilde A_n) = 1+ \sum_{i=1}^n m_i + \sum_{j=4}^n \sum_{i=2}^{j-2} m_j m_i + m_{1}\sum_{i=3}^{n-1} m_i + \cdots
\end{equation}
where we omitted products with more than two terms. We can rewrite it as 
\begin{eqnarray}
\nonumber\text{Tr}(\tilde A_n) &=& 1+ \sum_{i=1}^n m_i + \frac{1}{2}\left(\sum_{i=2}^n m_i\right)^2- \frac{1}{2}\sum_{i=2}^n m_i^2 - \sum_{i=3}^n m_i m_{i-1}\\
&&+ m_{1}\sum_{i=3}^{n-1} m_i + \cdots
\end{eqnarray}
and, therefore, for large $n$, we can write Eq. (\ref{TrAn}).
To evaluate the large $n$ limit of the trace, we  note that, if $m_n\neq 0$, 
\begin{equation}
\text{Tr}(\tilde A_{n-2}) = d^2_{n-2,1} + \frac{d^1_{n,1}}{m_{n}}
\end{equation}
and $d^2_{n,1}$ can be obtained from $d^1_{n,1}$ by permuting the site indices as follows $n\leftrightarrow 1$, $n-1 \leftrightarrow 2$, and so on. Thus we have to calculate the large-$n$ limit of $d^1_{n,1}$. Actually we have to determine when $\lim_{n\rightarrow\infty}d^1_{n,1}=0$. We get
\begin{equation}\label{eq dR}
d^1_{i,1} = \frac{m_i}{m_{i-1}}R_{i-2} d^1_{i-1,1}\,,
\end{equation}
where we have defined
\begin{equation}
R_{i} = \frac{1+\sum_{j=2}^i d^1_{j,1}}{1+ \sum_{j=2}^{i-1} d^1_{j,1}}\,,
\end{equation}
which fulfills Eq.~(\ref{recursive}), 
with $R_1=1$, 
and therefore, can be written as a random continued fraction, Eq. (\ref{continued}). 
Thus, we get $|\text{Tr}(\tilde A_n)|\sim \prod_{i=1}^n |R_i| = e^{n\langle \ln|R_i|\rangle}$ as $n \to \infty$, where the average can be evaluated using the probability distribution of the convergents $R_i$.

An alternative way to derive the trace $\text{Tr}(\tilde A_n)$ is the following. 
Performing the products we get
\begin{eqnarray}
\nonumber &&\text{Tr}(\tilde A_n)=(-1)^n\\
\nonumber &&\left\{m_2\Big[m_4\big(m_6(\cdots)\big)+m_5(\cdots)\Big]\right.
+m_3\big(m_5(\cdots)\big)+m_4(\cdots)\\
&&\left.+m_1\Big[m_3\big(m_5(\cdots)\big)+m_4(\cdots)\Big]\right\}
\label{Trm}
\end{eqnarray}
This expression can be obtained rewriting $A_i$ as
\be
A_i=B-m_i C\,,
\ee
where $B=\left(  \begin{array}{rr}
    -1 & 0 \\
    1 & 0 \\
  \end{array}
\right)$ and 
$C=\left(  \begin{array}{rr}
    0 & -1 \\
    0 & 0 \\
  \end{array}
\right)$.
The matrices $C$, $B$, $CB$ and $BC$ form a closed algebra and are such that
\be
BB=-B\,,\;
CC=0\,,\;
BCB=-B\,,\;
CBC=-C
\ee
and
\be
\text{Tr}(B)=\text{Tr}(CB)=\text{Tr}(BC)=-1,\;\;\text{Tr}(C)=0\,.
\ee
Making the product $A_iA_{i-1}$ we get $B\rightarrow -B$ or $B\rightarrow -m_i CB$,  and $CB\rightarrow -B$, while  $C\rightarrow BC$ and $BC\rightarrow -BC$ or $BC\rightarrow m_i C$. We can draw two graphs which are two trees of descendants with ancestors $B$ and $C$, the first generates only a cascade of $B$ and $CB$ while the latter generates $BC$ and $C$. 
The trace is, therefore, equal to the sum of all possible paths along the tree graphs, excluding those which end with $C$, since $\text{Tr}(C)=0$.
The number of these paths, after $n$ steps, are $F_n+2F_{n-1}$ where $F_n$ are the Fibonacci numbers. From Eq. (\ref{Trm}), or by counting the  paths as described above, we get that, after relabeling $i \leftrightarrow n-i$, we have to solve the following recursive equations
\be
\label{Fn}
{\cal F}_i={\cal F}_{i-1}+m_i{\cal F}_{i-2}
\ee
with initial conditions ${\cal F}_0=0$ ${\cal F}_1=1$, such that, given the solution ${\cal F}_n$, the trace of $\tilde A_n$, for large $n$, is 
\be
\text{Tr}(\tilde A_n)=(-1)^n\left({\cal F}_n+2 m\, {\cal F}_{n-1}\right)\,.
\ee
which can be also written as $\text{Tr}(\tilde A_n)=(-1)^n\left(2\,{\cal F}_{n+1}-{\cal F}_{n}\right)$. Notice that, for $m_i=1$, from Eq. (\ref{Fn}), we get ${\cal F}_{n}=F_n$, the Fibonacci numbers, then $\text{Tr}(\tilde A_n)=(-1)^n\left({F}_n+2 {F}_{n-1}\right)$. Generally, for integer $m_i=m$, the solution ${\cal F}_{n}$ of Eq. (\ref{Fn}) is a Lucas sequence, named $(1,m)$-Fibonacci sequence.\\
Defining the ratio
\be
R_i=\frac{{\cal F}_{i}}{{\cal F}_{i-1}}\,,
\ee
Eq. (\ref{Fn}) can be written as
$R_i=1+\frac{m_i}{R_{i-1}}$, which is Eq. (\ref{recursive}). As a result, $|\text{Tr}(\tilde A_n)|\simeq |{\cal F}_n|=\prod_{i=2}^n |R_i|$. 

\section{Case r=3}\label{app. r3}
Let us investigate the case with $r=3$, focusing on $\Delta_\ell=w_\ell=w$. It is easy to see that $D_n=\det(\tilde A_n)=\prod_{i=1}^n m_i$, therefore $|D_n| = e^{n \langle\ln |m_i|\rangle}$. To evaluate the trace $T_n=\textrm{Tr}(\tilde A_n)$ and $T'_n=\big((\text{Tr}(\tilde A_n))^2-\text{Tr}(\tilde A_n^2)\big)/2$, we consider $n$ as a multiple of 3, i.e., $n=3k$ with $k$ a positive integer. For $T_n$ we get
\begin{equation}
T_n = 2 + \sum_{i_e} p_n(i_e)
\end{equation}
where $p_n(i_e)$ are all the different products (without repetitions) of the lower endpoints of the non-crossing partitions of $(m_n,m_{n-1},\cdots,m_1)$ and cyclic permutations, e.g., $(m_{n-1},\cdots,m_1,m_n)$ and so on, with $3k$ elements with $k$ nonnegative integer. E.g., for $n=6$, we get $T_6 = 2 + \sum_{i=1}^6 m_i + m_1 m_4 + m_2 m_5 + m_3 m_6$, where $m_i$ comes from the partition $(m_{i-1},\cdots,m_1,m_6,\cdots,m_i)$, $m_1 m_4$ comes from the partition $(m_6,m_5,m_4)\cup (m_3,m_2,m_1)$, $m_2 m_5$ comes from the partition $(m_1,m_6,m_5)\cup (m_4,m_3,m_2)$ and $m_3 m_5$ comes from the partition $(m_2,m_1,m_6)\cup (m_5,m_4,m_3)$. 
Concerning $T'_n$, we get
\begin{equation}
T'_n = 1 + \sum_{i_e} p'_n(i_e)
\end{equation}
where $p'_n(i_e)$ are  all the different products of the lower endpoints of certain non-crossing partitions of $(m_n,m_{n-1},\cdots,m_1)$ and cyclic permutations, e.g., $(m_{n-1},\cdots,m_1,m_n)$ and so on. E.g., for $n=3$, we get $T'_3=1+\sum_{i=1}^3 m_i + m_1 m_2+ m_1 m_3 + m_2 m_3$.

We note that we can write $T_n$ and $T'_n$ in the forms $T_n = e^{n \langle \ln|K_l| \rangle}$ and $T'_n = e^{n \langle \ln|K'_l| \rangle}$, where $K_l$ and $K'_l$ are continued fractions uniquely determined by $(m_l,m_{l-1},\cdots,m_1)$. Let us consider $T_n$. We define $s_i$ with $i=1,\cdots,n$ the solutions of the $n$ equations
\begin{equation*}
2 + \sum_{i_e} p_l(i_e) = s_1(s_1+s_1 s_2)\cdots(s_1 +s_1s_2 +\cdots + s_1 s_2\cdots s_l)
\end{equation*}
for $l=1,\cdots, n$, with $p_l(i_e)$ as defined above. 
The terms  $K_l \equiv (s_1 + s_1 s_2 + \cdots +s_1 s_2 \cdots s_l)$ 
can be written as Euler's continued fractions,  
therefore, 
we can write $|T_n| = \prod_{l=1}^n |K_l| = e^{n \langle \ln|K_l| \rangle}$. An analogous form can be obtained for $T'_n$.
\section{Arbitrary $r$}\label{app. any r}
For an arbitrary $r$, for $w_\ell= \Delta_\ell$, we get the following generalized eigenvalue equation
\begin{equation}
\sum_{k=0}^r (-1)^k\lambda_n^{r-k} T^{(k)}_n = 0
\end{equation}
where $T^{(k)}_n = \text{Tr}(\bigwedge^k \tilde A_n)$  is the trace of the kth exterior power of $\tilde A_n$, which is defined by 
\begin{equation*}
T^{(k)}_n = \frac{1}{k!}\det\left(
  \begin{array}{ccccc}
    \text{Tr}(\tilde A_n) & k-1 & 0 & \cdots & 0 \\
    \text{Tr}(\tilde A_n^2) & \text{Tr}(\tilde A_n) & k-2 & \cdots & 0 \\
    \vdots & \vdots & \vdots & \ddots & \vdots \\
    \text{Tr}(\tilde A_n^{k-1}) & \text{Tr}(\tilde A_n^{k-2}) & \text{Tr}(\tilde A_n^{k-3}) & \cdots & 1 \\
    \text{Tr}(\tilde A_n^k) & \text{Tr}(\tilde A_n^{k-1}) & \text{Tr}(\tilde A_n^{k-2}) & \cdots & \text{Tr}(\tilde A_n) \\
  \end{array}
\right)
\end{equation*}
Then, it is easy to show that
\begin{eqnarray*}
 &&T_n^{(k)}\to 0, \forall k\in I_{1,r}  \Rightarrow \exists\,\text{r MZMs}\,,\\
\nonumber &&\left(T_n^{(k)} \to 0,   \forall k\in I_{2,r} \right)  \textrm{or} \,
\left(\frac{T_n^{(k)}}{T_n^{(1)}}\to 0, \forall k\in I_{2,r} \right) \Rightarrow \exists\,\text{r-1 MZMs}\,,\\
&& \qquad \vdots\\
&&T^{(r)}_n\to 0 \llor \frac{T^{(r)}_n}{T^{(1)}_n}\to 0 \;\;\textrm{or}\,\cdots\,\textrm{or}\;\;\frac{T^{(r)}_n}{T^{(r-1)}_n}\to 0 \Rightarrow \exists\,\text{1 MZM}\,.
\end{eqnarray*}
where $I_{i,j}=\{i,i+1,\cdots,j\}$. On the other hand, for $w_\ell\neq \Delta_\ell$, defining 
$T^{(k)}_n = \text{Tr}(\bigwedge^k \tilde {\cal A}_n)$, we get the eigenvalue equation
\begin{equation}\label{eq. app eg}
\sum_{k=0}^{2r} (-1)^k\lambda_n^{2r-k} T^{(k)}_n = 0
\end{equation}
In this case the conditions are modified as it follows
\begin{eqnarray*}
 &&T_n^{(k)}\to 0, \forall k\in I_{1,2r}  \Rightarrow v^< \hspace{-0.05cm}=2r\,,\\
&&\left(T_n^{(k)} \to 0, \forall k\in I_{2,2r} \right) \textrm{or}\, \left(\frac{T_n^{(k)}}{T_n^{(1)}}\to 0, \forall k\in I_{2,2r} \right) \Rightarrow v^<\hspace{-0.05cm}=2r-1\,,\\
\nonumber && \qquad \vdots\\
&&T^{(2r)}_n\to 0 \llor \frac{T^{(2r)}_n}{T^{(1)}_n}\to 0 \;\;\textrm{or}\,\cdots\,\textrm{or}\;\;
\frac{T^{(2r)}_n}{T^{(2r-1)}_n}\to 0 \Rightarrow v^< \hspace{-0.05cm}=1\,,
\end{eqnarray*}
otherwise $v^<=0$. After defining 
\be
{T'}^{(k)}_n=\frac{T^{(2r-k)}_n}{T^{(2r)}_n}
\ee
we get similar conditions for $v^>$, which read
\begin{eqnarray*}
 &&{T'}_n^{(k)}\to 0, \forall k\in I_{1,2r}  \Rightarrow v^>\hspace{-0.05cm}=2r\,,\\
&&\left({T'}_n^{(k)} \to 0, \forall k\in I_{2,2r} \right) \textrm{or}\, \left(\frac{{T'}_n^{(k)}}{{T'}_n^{(1)}}\to 0, \forall k\in I_{2,2r} \right) \Rightarrow v^>
\hspace{-0.1cm}=2r-1, \\
\nonumber && \qquad \vdots\\
&&{T'}^{(2r)}_n\to 0 \llor \frac{{T'}^{(2r)}_n}{{T'}^{(1)}_n}\to 0 \,\;\textrm{or}\,\cdots\,\textrm{or}\;\,
\frac{{T'}^{(2r)}_n}{{T'}^{(2r-1)}_n}\to 0 \Rightarrow v^>\hspace{-0.05cm}=1\,,
\end{eqnarray*}
otherwise $v^>=0$. %
We, then, obtain the number of MZMs which is given by $\max(v^<,v^>)-r$, where typically $v^>=2r-v^<$.

\section{Calculation of the self-energy} \label{app. s e}
We write $H_0=\sum_{m,n}c_m \mathcal{H}_{mn}c_n$, where $i \mathcal{H}$ is the real and skew-symmetric matrix
\begin{eqnarray}
\nonumber \mathcal{H} &=& \sum_{i,j} \ket{i}\bra{j}\otimes \mathcal{H}_{i,j} = \sum_j \ket{j}\bra{j} \otimes  \mathcal{H}_0  \\
&& + \sum_j\sum_\ell  \ket{j} \bra{j+\ell} \otimes \mathcal{H}_\ell +  \ket{j+\ell}\bra{j}\otimes (\mathcal{H}_\ell)^\dagger\,,
\end{eqnarray}
where $\mathcal{H}_0$ and $\mathcal{H}_\ell$ are the matrices $\mathcal{H}_0 = \mu \tau_2 /4$ and $\mathcal{H}_\ell = w \ell^{-\alpha}\tau_2/4 + i \Delta \ell^{-\beta} \tau_1/4$, where the nth component of $\ket{i}$ is $(\ket{i})_n=\delta_{n,i}$.
We consider periodic boundary conditions, and we change the basis by defining the vectors $\ket{k}$ such that $\ket{j} = \sum_k e^{-i k j}\ket{k}/\sqrt{L}$.  We can then write  $\mathcal{H} = \sum_k \ket{k} \bra{k}\otimes \mathcal{H}(k)$, where %
$\mathcal{H}(k) = \big( (\mu + w g(k))\tau_2 - \Delta f(k)\tau_1\big)/4$. The inverse  of the matrix $\mathcal{H}$ reads
\begin{equation}\label{inverse}
(\mathcal{H})^{-1} = -4 i \sum_k \ket{k}\bra{k}\otimes \left(
                         \begin{array}{cc}
                          0  & 1/X_+ \\
                           -1/X_- &  0\\
                         \end{array}
                       \right) \,,
\end{equation}
where $X_\pm = \mu + w g(k)\pm i \Delta f(k)$.
We now add a disorder term $H_1=-\sum_{j=1}^L \omega_j n_j$ to the Hamiltonian $H_0$, where $\omega_j=\mu_j-\mu$, corresponding to the matrix $\mathcal H_1 = \sum_i\omega_i \ket{i}\bra{i}\otimes \tau_2 /4$. For $E=0$, we get $\mathcal{G}_0=-(\mathcal{H})^{-1} $, and the self-energy $\Sigma' \approx \langle \mathcal{H}_1 \mathcal{G}_0 \mathcal{H}_1 \rangle$ reads
\begin{equation}
\Sigma' \approx  \frac{\delta \mu}{4} \sum_{j=1}^L \ket{j}\bra{j}\otimes \tau_2  + \sum_j\sum_\ell  \ket{j} \bra{j+\ell} \otimes \Sigma'_\ell +  \ket{j+\ell}\bra{j}\otimes (\Sigma'_\ell)^\dagger
\end{equation}
with $\Sigma'_\ell = \delta w_\ell\tau_2/4 + i \delta \Delta_\ell \tau_1/4$, where 
$\delta \mu$, $ \delta w_\ell$ and $ \delta \Delta_\ell$ given by Eqs. (\ref{dmu}), (\ref{dwl}) and (\ref{dDl}), 
from which we get the self-energy $\Sigma$ in terms of Majorana operators reported in Eq.~\eqref{eq. se s}.

\end{document}